\documentclass[a4paper,11pt]{article}
\pdfoutput=1 

\usepackage{jheppub} 

\usepackage[T1]{fontenc} 
\usepackage{comment}

\newcommand{\be}{\begin{equation}}
	\newcommand{\ee}{\end{equation}}

\def\l{{\lambda}}

\newcommand{\bea}{\begin{eqnarray}}
	\newcommand{\eea}{\end{eqnarray}}
\newcommand{\ba}{\begin{eqnarray}}
	\newcommand{\ea}{\end{eqnarray}}

\title{Inconsistency of point-particle dynamics on  higher-spin backgrounds: massive particles}

\author[a,b]{Vyacheslav Ivanovskiy}
\author[b,c]{and Dmitry Ponomarev}

\affiliation[a]{Moscow Institute of Physics and Technology,
  Dolgoprudny, 141701, Russia}
\affiliation[b]{Institute for Theoretical and Mathematical Physics,\\
Lomonosov Moscow State University,  Moscow, 119991, Russia}
\affiliation[c]{I.E. Tamm Theory Department, Lebedev Physical Institute,
 Moscow, 119991, Russia}

\emailAdd{ivanovskiy.va@phystech.edu}
\emailAdd{ponomarev@lpi.ru}

\abstract{Previously, we showed that massless scalar point particles cannot propagate on classical backgrounds of chiral higher-spin theories. This conclusion was derived from the analysis of the light-cone consistency conditions occurring at the second order in interactions. 
In the present paper, we extend this result to the case of massive particles, showing that these cannot propagate on chiral higher-spin backgrounds either. 
In order to do that, we use a different and  more direct approach, which does not rely on special simplifications occurring for massless particles. 
Namely, we solve the light-cone consistency conditions at the given order in complete generality and then 
show that all the Hamiltonians found are inevitably non-local. We emphasise connections between  the resulting procedure and the on-shell methods applied to worldline scattering observables. 
 }

\begin{document} 
\maketitle
\flushbottom

\section{Introduction}

When constructing interactions of massless fields it can be beneficial to use light-cone gauge. By going to light-cone gauge one deals with the physical degrees of freedom only, which allows one to avoid difficulties with maintaining gauge invariance at non-linear level. This simplification, however, comes at a cost: light-cone gauge breaks manifest Lorentz invariance of covariant descriptions, therefore, maintaining it at the non-linear level requires additional efforts. The standard way to control Lorentz invariance in this case is to use the approach of \cite{Dirac:1949cp}, which amounts to the construction of  phase space charges associated with Lorentz generators and ensuring that these commute properly. This approach towards the construction of interactions in massless theories is known as the light-cone formalism. It is used rather extensively in the higher-spin literature. In particular, first interactions of massless higher-spin fields were obtained employing the light-cone formalism \cite{Bengtsson:1983pg,Bengtsson:1986kh}. Since then it was actively used and applied to different setups, not limited to interactions of massless fields,
see e.g. \cite{Metsaev:1993ap,Metsaev:2005ar,Metsaev:2007rn,Metsaev:2017cuz,Metsaev:2018xip,Metsaev:2025qkr} and references therein.

Despite the ideas behind the light-cone formalism are conceptually simple, the consistency conditions resulting from it are rather complicated. Moreover, these consistency conditions become increasingly more complicated as the order of interactions grows. At the leading order in interactions the light-cone consistency conditions were analysed for different theories, see e. g. \cite{Bengtsson:1983pg,Bengtsson:1986kh,Metsaev:1993ap,Metsaev:2005ar,Metsaev:2007rn,Metsaev:2017cuz,Metsaev:2018xip,Metsaev:2025qkr}, resulting in complete classifications of consistent interactions in the respective cases. At the same time, the literature on the light-cone deformation procedure beyond the leading order is very limited and our understanding of how to approach the associated consistency conditions systematically and efficiently is lacking. This lack of understanding is especially pressing, considering that the stringent constraints responsible for deciding whether  one or another interacting theory exists occur at higher orders -- typically, at the second order in the coupling constant -- of the perturbative analysis.

Important progress in this direction was achieved in \cite{Metsaev:1991mt,Metsaev:1991nb}, where interactions of massless higher-spin fields in four-dimensional flat space were studied at the second order in the coupling constant.  In these works it was observed that the light-cone consistency conditions in the given case contain two simple sectors, which do not involve contributions from higher-order vertices, thus, leading to a pair of constraints featuring only cubic coupling constants as unknowns. The resulting constraints turned out to be simple enough to be solved systematically. As a result, a closed formula expressing the cubic coupling constants in terms of two independent parameters was found. Moreover, this analysis straightforwardly entails chiral higher-spin theories explicitly suggested in \cite{Ponomarev:2016lrm}.
The same approach was also used recently \cite{Ponomarev:2024jyg} to construct families of chiral higher-spin theories of a more general type as well as to classify various contractions and truncations \cite{Serrani:2025owx} of chiral higher-spin theories.

The aforementioned sectors of the light-cone consistency conditions can be referred to as the self-dual and anti self-dual ones, 
because these give the complete consistency conditions for self-dual and anti self-dual theories respectively. Once both the self-dual and anti self-dual consistency conditions are solved, one  still has to ensure that the consistency conditions are satisfied in the remaining, mixed, sector, which features cubic vertices of both chiralities.
The consistency conditions in the mixed sector do involve higher-order vertices non-trivially, which entails  genuine difficulties associated with the analysis of the light-cone deformation procedure at the given order.

The second-order light-cone consistency conditions for massless fields in four-di\-men\-sion\-al Minkowski space in the mixed sector were also solved in \cite{Metsaev:1991nb}.
This result is based on guessing a particular solution for the quartic light-cone Hamiltonian, which is given by a singular exchange-like expression.
The general solution is then given as the sum of this particular solution and the general solution of the homogeneous light-cone consistency conditions. 

This approach was further developed and extended to all orders in perturbations in \cite{Ponomarev:2016cwi}, where it was shown that  the light-cone consistency conditions for massless fields in four-dimensional Minkowski space are equivalent to the requirement that the off-shell amplitudes constructed from the light-cone action via the natural Feynman rules satisfy a version of the Ward identities.
Moreover, it was found that the general solution of these Ward identities is given by a certain off-shell extension of the spinor-helicity amplitudes.
The relevance of the spinor-helicity formalism in the context of the light-cone deformation procedure was also highlighted in earlier works \cite{Ananth:2012un,Akshay:2014qea,Bengtsson:2016jfk,Bengtsson:2016alt}.

 The result of \cite{Ponomarev:2016cwi} provides an efficient approach towards finding the general solution of the light-cone consistency conditions for interactions of massless higher-spin fields at all orders.  Namely, consistent light-cone vertices can be generated from any  set of spinor-helicity amplitudes just by inverting the Feynman rules. If, in addition, locality is required, it can be easily controlled at the level of amplitudes using the standard approaches, such as the on-shell methods, see e. g. \cite{Britto:2004ap,Britto:2005fq,Benincasa:2007xk,Benincasa:2011pg,Arkani-Hamed:2017jhn}. In other words, for massless four-dimensional theories \cite{Ponomarev:2016cwi} allows one to avoid the tedious  light-cone analysis by replacing it with more efficient manipulations with spinor-helicity amplitudes.

In the present paper we will study the consistency conditions for interactions of a massive scalar  point particle with background fields of chiral higher-spin theories at the second order in the coupling constant. In the previous paper \cite{Ivanovskiy:2025kok} we showed that massless particles  cannot interact with background chiral higher-spin fields. The analysis of the massless case was facilitated by simplifications analogous to those occurring for massless fields in the self-dual and anti self-dual sectors. These simplifications do not occur for interactions of massive point particles.

Our goal in the present paper is to complete the analysis of \cite{Ivanovskiy:2025kok} by showing that massive particles cannot propagate on chiral higher-spin backgrounds either. The problem of  consistent propagation of point particles on massless higher-spin backgrounds is, in particular, motivated by the attempts to construct a higher-spin extension of Riemannian geometry, see  \cite{Segal:2002gd,Ponomarev:2013mqa,Tomasiello:2024jyu}. Further details on our motivations and a more comprehensive list of references on recent developments in chiral higher-spin theories can be found in \cite{Ivanovskiy:2023aay,Ivanovskiy:2025kok}. 
The present paper is also aimed at developing a better understanding of systematics of the light-cone formalism at higher orders in the coupling constant.
 We find that the light-cone consistency conditions for the problem in question can be solved systematically and the worldline scattering observables naturally occur in this process. 
In other words, we find that the result of \cite{Ponomarev:2016cwi} naturally extends to worldline theories and provides an efficient approach towards solving the light-cone consistency conditions in the given case.

This paper is organised as follows. We start by reviewing the light-cone deformation procedure  in section \ref{sec:2}. It also contains a review of the leading order results on interactions of point particles with massless higher-spin fields as well as the necessary details on chiral higher-spin theories.
Then, in section \ref{sec:3} we present the light-cone consistency conditions on interactions between scalar point particles and chiral higher-spin fields occurring  at the second order in the coupling constant. Next, in section \ref{sec:4} we give their general solution. In the following section \ref{sec:5} we show that the general solution inevitably features singularities, thus, no local interactions of point particles with chiral higher-spin theories exist. 
Then, in section \ref{sec:6} we show how the previous discussion can be interpreted in terms of scattering amplitudes in the classical limit. Finally, we conclude in section \ref{sec:7}.

\section{Derivation of the constraints}
\label{sec:2}

The light-cone consistency conditions that we are to solve were already derived in \cite{Ivanovskiy:2025kok}. We will solve them in the following sections. In the present section we will briefly review how these were derived. We refer the reader interested in further details to \cite{Ivanovskiy:2025kok}.

According to Noether's theorem, a Poincare invariant theory has a set of conserved quantities $\mathcal Q [O]$ each corresponding to the associated Poincare algebra generator $O$. These charges, once written in terms of phase space variables, generate the phase space Poincare algebra action via the  Poisson (Dirac) bracket
\begin{equation}
\label{17apr1}
\delta_O f = [f,\mathcal Q [O]],
\end{equation}
where $f$ is an arbitrary function of phase space variables. Consistency of this phase space representation of the Poincare algebra with its commutation relations entails
\begin{equation}
\label{17apr2}
[\delta_{O^{(1)}},\delta_{O^{(2)}}]f = \delta_{[O^{(1)},O^{(2)}]} f, 
\end{equation}
where $O^{(1)}$ and $O^{(2)}$ is a pair of Poincare algebra generators, while the bracket on the right-hand side is the Lie algebra commutator. Combining (\ref{17apr1}), (\ref{17apr2}) with the Jacobi identity for the Poisson (Dirac) bracket, one finds 
\begin{equation}
\label{17apr3}
\big[\mathcal Q [ O^{(1)}], \mathcal Q [ O^{(2)}] \big] = \mathcal Q \big[ [O^{(1)},O^{(2)}]\big].
\end{equation}
Thus, every Poincare invariant theory is associated with a set of phase space charges, that satisfy (\ref{17apr3}). Vice versa, once a set of charges that satisfy (\ref{17apr3}) is available, one can pick a convenient time variable, identify the associated charge with the Hamiltonian and then,
employing the standard formula,
 construct the action of a Poincare invariant theory.

In the light-cone formalism one starts from a Poincare invariant free theory and then adds interactions perturbatively while maintaining Poincare invariance through the consistency condition (\ref{17apr3}). For the problem in question, phase space charges have the form
\begin{equation}
\label{17apr4}
\mathcal Q [O] = Q[O] + q[O],
\end{equation}
where the components $Q[O]$ and $q[O]$ are the field theory and the point particle contributions respectively. By substituting (\ref{17apr4}) into (\ref{17apr3}) and denoting 
\begin{equation}
\label{17apr6}
O^{(3)} \equiv [O^{(1)} ,O^{(2)} ], \qquad Q^{(i)} \equiv Q[O^{(i)}], \qquad  q^{(i)} \equiv q[O^{(i)}],
\end{equation}
we find 
\begin{equation}
		[q^{(1)},q^{(2)}]_p +[q^{(1)},Q^{(2)}]_\Phi + [Q^{(1)},q^{(2)}]_\Phi = q^{(3)}.
		\label{3.2}
	\end{equation}
Here $[\cdot, \cdot]_\Phi$ and $[\cdot, \cdot]_p$ denote the field theory and the point particle Poisson (Dirac) brackets respectively.  To arrive at (\ref{3.2}) we used that the field theory is Poincare invariant on its own. Besides that, we dropped the back-reaction terms, which correspond to divergent interactions of a point particle with the fields it sources. A detailed discussion of back-reaction terms can be found in \cite{Ivanovskiy:2023aay}.

The perturbative expansion of the Poincare charges on the field theory side has the form
\begin{equation}
\label{17apr5x1}
Q[O] = Q_2[O]+Q_3[O].
\end{equation}
Here $Q_2 [O]$ represents the free field theory charge, which is quadratic in fields. Next, $Q_3[O]$ is a field theory contribution, which is cubic in fields. In (\ref{17apr5x1}) we did not write higher-order field theory contributions, because the theories we are going to deal with have only cubic vertices. In turn, the point particle charge expansion 
\begin{equation}
\label{17apr5x2}
q[O]=q_0[O]+ q_1[O]+q_2[O]+\dots 
\end{equation}
 starts with a field-independent term $q_0[O]$, which corresponds to the free point particle theory. It is then followed by non-linear corrections $q_1[O]$, $q_2[O]$ and so on, where the lower label denotes the power with which fields contribute to the charge. 

In the Hamiltonian formalism the space and time variables are treated differently and, as a result, upon adding interactions, only the charges that involve time translations receive corrections \cite{Dirac:1949cp}. These charges are called dynamical. The remaining charges do not receive non-linear corrections and are referred to as kinematical. 
By denoting dynamical and kinematical charges as $D$ and $K$ respectively, one, schematically, has
\begin{equation}
\label{17apr7}
D=D_2 +\delta D, \qquad K=K_2.
\end{equation}
Accordingly, commutators of the Poincare algebra split into classes depending on the types of charges they feature. The first type of commutators is given by
\begin{equation}
\label{17apr8}
[K,K]=K,
\end{equation}
which are trivially satisfied at the non-linear level as a consequence of Poincare invariance of the free theory. The next class of commutators is of the form
\begin{equation}
\label{17apr9}
[K,D]=K, \qquad [K,D]=D
\end{equation}
and it is called kinematical. 
These entail linear constraints on the deformations of the dynamical charges
\begin{equation}
\label{17apr10}
[K_2,\delta D]=K_2, \qquad [K_2,\delta D]=\delta D.
\end{equation}
The fact that these constraints are linear in $\delta D$ makes them relatively easy to solve at all orders in fields. The final class of commutators has the schematic form
\begin{equation}
\label{17apr11}
[D,D]=0.
\end{equation}
Being quadratic in deformations $\delta D$, these constraints -- referred to as dynamical -- constitute the main difficulty of the light-cone deformation procedure. 

It is convenient to use the light-cone coordinates
\begin{equation}
\begin{split}
	x^+ \,=& \frac{1}{\sqrt{2}}(x^3+x^0),\qquad   x^-\, = \frac{1}{\sqrt{2}}(x^3-x^0),\\
	\label{29sep1}
	x \,= & \frac{1}{\sqrt{2}}(x^1-ix^2),\qquad   \bar x \,= \frac{1}{\sqrt{2}}(x^1+ix^2),
	\end{split}
\end{equation}
where $x^+$ is regarded as the time coordinate, while $x^{\perp} =\{x^-,x,\bar x \}$ are treated as spatial coordinates. With this choice of time, there are three dynamical generators
\begin{equation}
\label{17apr12}
P^-, \qquad J^{x-}, \qquad  J^{\bar x -}
\end{equation}
and three dynamical commutators 
\begin{equation}
\label{17apr13}
[P^-,J^{x-}]=0, \qquad [P^-,J^{\bar x-}]=0, \qquad [J^{x-},J^{\bar x -}]=0.
\end{equation}
It can be shown, see e. g. discussions in \cite{Ponomarev:2016lrm}, that once the constraints associated with the first two commutators are satisfied, the last constraint is satisfied automatically. Thus, at every order in perturbation theory we need to solve only two dynamical constraints, which are, moreover, related by complex conjugation. 

We have finished the discussion of the general structure of the light-cone consistency conditions relevant for the interactions between a point particle and a field theory. Below we will summarise the necessary results on the field theory side, as well as review the lower-order results on the point particle charges, which we obtained previously.

\subsection{Field theory generators}

We are going to deal with massless fields in Minkowski space. Dynamical charges at the free level are 
	\begin{equation}
	\begin{split}
		\label{1.15}
		Q_2[P^-] &= \sum_{\lambda}\int d^3x^\perp \partial^+\Phi^{-\lambda}  \frac{\partial \bar\partial}{\partial^+} \Phi^\lambda, \\
		  Q_2[J^{x-}] &= \sum_{\lambda}\int d^3x^\perp \partial^+\Phi^{-\lambda} \left(x\frac{\partial \bar\partial}{\partial^+} + x^- \partial- \lambda\frac{\partial}{\partial^+} \right) \Phi^\lambda,\\
		  Q_2[J^{\bar x-}] &= \sum_{\lambda}\int d^3x^\perp \partial^+\Phi^{-\lambda} \left(\bar x\frac{\partial \bar\partial}{\partial^+} + x^- \bar\partial
		+\lambda\frac{\bar\partial}{\partial^+}\right) \Phi^\lambda,
		  \end{split}
	\end{equation}
where 
\begin{equation}
\label{18apr1}
\partial \equiv \partial^x= \eta^{x\bar x}\partial_{\bar x} = \partial_{\bar x}, \qquad \bar\partial \equiv \partial^{\bar x} = \eta^{\bar x x}\partial_{x}= \partial_x
\end{equation}
and $\lambda$ is the helicity of the field. The field theory Dirac bracket is given by
	\begin{equation}
		\label{1.9}
		[\partial^+ \Phi^\lambda(x^\perp,x^+),\Phi^{\mu}(y^\perp,x^+)]_\Phi= \frac{1}{2}\delta^{\lambda+\mu,0}\delta^{3}(x^\perp,y^\perp).
	\end{equation}

We will focus on the coupling of a point particle to three theories: the chiral higher-spin theory, the Poisson chiral higher-spin theory and self-dual gravity. 
In all these three cases theories feature only cubic vertices. The associated dynamical charges can be universally written as
\begin{equation}
\label{19feb1}
\begin{split}
	&Q_3[P^{-}]=\sum\limits_{\l_1, \l_2,\l_3}\int d^{3+9}x^{\perp} \prod_{i=1}^3\delta^3(x^\perp-x^\perp_i) h_{\Phi}^{\l_1\l_2\l_3}
	\prod_{i=1}^3\Phi^{\lambda_i}(x^\perp_i),\\
	&Q_3[J^{x-}]=\sum\limits_{\l_1, \l_2,\l_3}\int d^{3+9}x^{\perp} \prod_{i=1}^3\delta^3(x^\perp-x^\perp_i) 
	\left(j^{\l_1\l_2\l_3}_{\Phi}+xh^{\l_1\l_2\l_3}_{\Phi}\right)\prod_{i=1}^3\Phi^{\lambda_i}(x^\perp_i),\\
	&Q_3[J^{\bar x-}]=\sum\limits_{\l_1, \l_2,\l_3}\int d^{3+9}x^{\perp} \prod_{i=1}^3\delta^3(x^\perp-x^\perp_i) 
	\left(\bar j^{\l_1\l_2\l_3}_{\Phi}+\bar xh^{\l_1\l_2\l_3}_{\Phi}\right)\prod_{i=1}^3\Phi^{\lambda_i}(x^\perp_i),
	\end{split}
\end{equation}
where 
\begin{equation}
\label{19feb1x1}
\begin{split}
	h_{\Phi}^{\l_1\l_2\l_3}&={C^{\l_1\l_2\l_3}}
	\frac{
		\bar{\mathbb{P}}
		^{\l_1+\l_2+\l_3}}
	{\partial^{+\l_1}_1\partial^{+\l_2}_2\partial^{+\l_3}_3},\\
	j^{\l_1\l_2\l_3}_{\Phi}&=
	\frac{2}{3} {C^{\l_1\l_2\l_3}} \left(\partial^{+}_1(\lambda_2-\lambda_3)+\partial^{+}_2(\lambda_3-\lambda_1)+\partial^{+}_3(\lambda_1-\lambda_2)\right)
	\frac{\bar{\mathbb{P}}^{\l_1+\l_2+\l_3-1}}
	{\partial^{+\l_1}_1\partial^{+\l_2}_2\partial^{+\l_3}_3},\\
	\bar j^{\l_1\l_2\l_3}_{\Phi}&=0	
	\end{split}
\end{equation}
and 
\begin{equation}
\label{28feb2}
\bar{\mathbb{P}}\equiv\frac{1}{3} \big(\partial_{x_3}(\partial^{+}_1-\partial^+_{2})+\partial_{x_1}(\partial^{+}_2-\partial^+_{3})+\partial_{x_2}(\partial^{+}_3-\partial^+_{1})\big).
\end{equation}

Coefficients $C^{\l_1\l_2\l_3}$ are the coupling constants. For self-dual gravity only helicities $+2$ and $-2$ are present in the spectrum, while the only non-trivial coupling constants are 
\begin{equation}
\label{28feb3}
C^{-2, 2, 2}=C^{2, -2, 2}=C^{2, 2, -2}=l,
\end{equation}
where $l$ is a parameter of dimension length, which is related to Newtons' coupling constant. For the chiral higher-spin theory fields of all helicities are present and the coupling constants are 
\begin{equation}
\label{28feb4}
C^{\l_1\l_2\l_3}=\frac{l^{\l_1+\l_2+\l_3-1}}{\Gamma(\l_1+\l_2+\l_3)}.
\end{equation}
Finally, for the Poisson chiral higher-spin theory the coupling constants read
\begin{equation}
\label{28feb5}
C^{\l_1\l_2\l_3}=2l\delta(\l_1+\l_2+\l_3-2).
\end{equation}

\subsection{Lower order point particle generators}

Here we review some intermediate results from the lower-order analysis of interactions between point particles and massless higher-spin fields, that will be relevant in the following. More details can be found in  \cite{Ivanovskiy:2023aay,Ivanovskiy:2025kok}. 

A scalar point particle freely propagating in Minkowski space has the following dynamical charges
\begin{equation}
\label{18apr2}
\begin{split}
q_0[P^-] = H_p,\qquad 
q_0[J^{x-}] =H_p x+ p^x x^-, \qquad  
q_0[J^{\bar x-}] =H_p \bar x+ p^{\bar x} x^-,
\end{split}
\end{equation}
where 
\begin{equation}
\label{18apr3}
H_p \equiv \frac{p_x p_{\bar x}}{p_-}+\frac{m^2}{2p_-}.
\end{equation}
Point particle momenta $p^\perp = \{p^+, p^x,p^{\bar x} \}$ and spatial coordinates have the standard Poisson bracket
	\begin{equation}
		[f, g]_p\equiv \frac{\partial f}{\partial x^i}\frac{\partial g}{\partial p_i}-\frac{\partial f}{\partial p_i}\frac{\partial g}{\partial x^i},
	\end{equation}
	where $i$ runs over spatial indices only.

The most general ansatz for the deformation of the point particle charges at the $n$-th order in fields is	
	\begin{equation}
		\label{3.18}
	\begin{split}
	&q_n[P^-]=\sum\limits_{\lambda_i}\int d^{3n}z^\perp
	\left(\prod\limits_{i=1}^n\delta^3(x^\perp-z_i^{\perp})\right) h^{\lambda_1 \dots \lambda_n} \prod\limits_{i=1}^n \Phi^{\lambda_i}(z_i^{\perp}),
	\\
	&q_n[J^{x-}]=\sum\limits_{\lambda_i}\int d^{3n}z^\perp \left(\prod\limits_{i=1}^n\delta^3(x^\perp-z_i^{\perp})\right)
	 \left(
	j^{\lambda_1 \dots \lambda_n}
	+x \ h^{\lambda_1 \dots \lambda_n}\right)
	\prod\limits_{i=1}^n \Phi^{\lambda_i}(z_i^{\perp}),\\
	&q_n[J^{\bar x -}]=\sum\limits_{\lambda_i}\int d^{3n}z^\perp \left(\prod\limits_{i=1}^n\delta^3(x^\perp-z_i^{\perp})\right)\left(
	\bar j^{\lambda_1 \dots \lambda_n}
	+\bar x \ h^{\lambda_1 \dots \lambda_n}\right) \prod\limits_{i=1}^n \Phi^{\lambda_i}(z_i^{\perp}).
	\end{split}
\end{equation}
	Here $x$ is the point particle coordinate, while $z_i$'s are the coordinates of fields. The presence of the delta functions ensures that the particle interacts with the values of fields and their derivatives at the point where the particle is located. Functions $h^{\lambda_1 \dots, \lambda_n}$, $j^{\lambda_1 \dots \lambda_n}$ and $\bar j^{\lambda_1 \dots \lambda_n}$ are arbitrary functions of point particle phase space variables  $x^\perp$ and $p^\perp$ as well as of derivatives $\partial^\perp_{z_i}$ acting on fields. 

By solving the kinematical constraints, which can be done straightforwardly at all orders, one finds that 	
\begin{equation}
\label{4.17}
\begin{split}
	h^{\lambda_1 \dots \lambda_n}&=\frac{A^{\lambda_1 \dots \lambda_n}(\sigma_{z_i}, \sigma_{\bar z_i}, s_i)}{p^+}, \\
		j^{\lambda_1 \dots \lambda_n}&=\frac{a^{\lambda_1 \dots \lambda_n}(\sigma_{z_i}, \sigma_{\bar z_i}, s_i)}{p^+},  \\
			\bar j^{\lambda_1 \dots \lambda_n}&=\frac{\bar a^{\lambda_1 \dots \lambda_n}(\sigma_{z_i}, \sigma_{\bar z_i}, s_i)}{p^+},
		\end{split}
\end{equation}	
	where 
	\begin{equation}
	\sigma_{ z_i}=p_x-\partial_{z_i} \frac{p^+}{\partial^+_i}, \qquad \sigma_{ \bar z_i}=p_{\bar x}-\partial_{\bar z_i} \frac{p^+}{\partial^+_i}, \qquad 
	s_i\equiv \frac{\partial^+_i}{p^+}
	\label{4.018}
\end{equation}
and $A^{\lambda_1 \dots \lambda_n}$, $a^{\lambda_1 \dots \lambda_n}$ and $\bar{a}^{\lambda_1 \dots \lambda_n}$  are arbitrary functions of their arguments, satisfying homogeneity conditions
\begin{equation}
\begin{split}
\label{21apr1x1}
	\sum_{i=1}^n \left(\sigma_{z_i}\frac{\partial A^{\lambda_1 \dots \lambda_n}}{\partial \sigma_{z_i}}-\sigma_{\bar z_i}\frac{\partial A^{\lambda_1 \dots \lambda_n}}{\partial \sigma_{\bar z_i}}-\lambda_i  A^{\lambda_1 \dots \lambda_n}\right)=0,\\
	\sum_{i=1}^n \left(\sigma_{z_i}\frac{\partial a^{\lambda_1 \dots \lambda_n}}{\partial \sigma_{z_i}}-\sigma_{\bar z_i}\frac{\partial a^{\lambda_1 \dots \lambda_n}}{\partial \sigma_{\bar z_i}}-\lambda_i  a^{\lambda_1 \dots \lambda_n}\right)+a^{\lambda_1 \dots \lambda_n}=0,\\
	\sum_{i=1}^n \left(\sigma_{z_i}\frac{\partial \bar a^{\lambda_1 \dots \lambda_n}}{\partial \sigma_{z_i}}-\sigma_{\bar z_i}\frac{\partial \bar a^{\lambda_1 \dots \lambda_n}}{\partial \sigma_{\bar z_i}}-\lambda_i  \bar a^{\lambda_1 \dots \lambda_n}\right)-\bar a^{\lambda_1 \dots \lambda_n}=0.
	\end{split}
\end{equation}

Next, we proceed to the dynamical constraints. At the leading order in fields (\ref{3.2}) acquires the form
\begin{equation}
	\label{2.6}
	\begin{split}
	[Q_2^{(1)},q_1^{(2)}]_\Phi+[q_1^{(1)},Q_2^{(2)}]_\Phi+
	[q_0^{(1)},q_1^{(2)}]_p
	 +[q_1^{(1)},q_0^{(2)}]_p
	=q_1^{(3)}.
	\end{split}
\end{equation}
For example, the dynamical constraint associated with the commutator $[P^-,J^{x-}]=0$ at the leading order gives
\begin{equation}
	\label{4.25}
	[Q_2[P^-],q_1[J^{x-}]]_\Phi+[q_1[P^-],Q_2[J^{x-}]]_\Phi+
	[q_0[P^-],q_1[J^{x-}]]_p +[q_1[P^-],q_0[J^{x-}]]_p
	=0.
\end{equation}
The constraint that follows from $[P^-,J^{\bar x-}]=0$ is analogous. By solving these, one finds 
\begin{equation}
	\label{4.1}
	\begin{split}
		&h^{\l}= \left(C^{\lambda}\frac{\sigma^{\lambda}_{z_1}}{p^+}+\bar C^{-\lambda}\frac{\sigma^{-\lambda}_{\bar{z}_1}}{p^+}\right),\\
		&j^{\l}=\l C^{\l}\frac{\sigma^{\lambda -1}_{z_1}}{s_{1} p^+}, \qquad 	\bar j^{\l}=-\l \bar C^{-\l}\frac{\sigma^{-\lambda -1}_{\bar{z}_1}}{s_{1} p^+},
		\end{split}
	\end{equation}
	where $C^\lambda$ and $\bar C^{\lambda}$ are arbitrary coupling constants satisfying
	\begin{equation}
		\label{13jun4}
		C^{\lambda}=0, \qquad \lambda<0 \qquad \text{and} \qquad \bar C^{-\lambda}=0, \qquad \lambda \ge 0. 
	\end{equation}

Before proceeding to the second-order consistency conditions and their analysis, let us note that the absence of consistent interaction between massless point particles and chiral higher-spin fields strongly suggests, that massive point particles cannot interact with these either. Indeed, if consistent interactions of massive particles with chiral higher-spin fields did exist and, moreover, the coupling of a point particle to the scalar field at the leading order was non-trivial, then, by shifting the scalar field by the appropriate constant one would have found 
a consistent interacting theory of a massless point particle with chiral higher-spin fields. This can be seen by noticing that the leading order correction to the Hamiltonian (\ref{4.1})
for $\lambda=0$ and for constant $\Phi^0$ has the same form as the mass term in the free Hamiltonian (\ref{18apr3}). However, as shown  in \cite{Ivanovskiy:2025kok}, massless particles cannot interact with chiral higher-spin theories, so the same should apply for massive particles as well. 
 This argument, however, relies on the assumption that the leading-order vertex for the point particle interaction with the scalar field is non-trivial. Below we will carry out a comprehensive analysis of the light-cone consistency conditions which does not rely on this assumption.

\section{Second-order consistency conditions}
\label{sec:3}

In this section we proceed with the analysis of the dynamical constraints at the second order in the coupling constant. 
At this order (\ref{3.2}) gives
\begin{equation}
	\label{2.6x1}
	\begin{split}
	&[Q_2^{(1)},q_2^{(2)}]_\Phi +	[Q_3^{(1)},q_1^{(2)}]_\Phi
	+[q_2^{(1)},Q_2^{(2)}]_\Phi+[q_1^{(1)},Q_3^{(2)}]_\Phi\\
	&\qquad \qquad \qquad +
	[q_0^{(1)},q_2^{(2)}]_p +[q_1^{(1)},q_1^{(2)}]_p
	 +[q_2^{(1)},q_0^{(2)}]_p
	=q_2^{(3)}.
	\end{split}
\end{equation}
In particular, the constraint associated with the commutator $[P^-,J^{x-}]=0$ results in
\begin{equation}
	\label{4.25x1}
	\begin{split}
	&[Q_2[P^-],q_2[J^{x-}]]_\Phi+[Q_3[P^-],q_1[J^{x-}]]_\Phi+
	[q_2[P^-],Q_2[J^{x-}]]_\Phi+[q_1[P^-],Q_3[J^{x-}]]_\Phi\\
	&\qquad\qquad \qquad\qquad +
	[q_0[P^-],q_2[J^{x-}]]_p + [q_1[P^-],q_1[J^{x-}]]_p +[q_2[P^-],q_0[J^{x-}]]_p
	=0.
	\end{split}
\end{equation}
The constraint associated with $[P^-,J^{\bar{x}-}]=0$ can be easily obtained by replacing $J^{x-}$ in (\ref{4.25x1}) with $J^{\bar{x}-}$.

Equation (\ref{4.25x1}) involves two unknown functions, which enter linearly. These are $A^{\lambda_1\lambda_2}$ and $a^{\lambda_1\lambda_2}$, defining the second-order charges 
$q_2[P^-]$ and $q_2[J^{x-}]$, see (\ref{3.18}) and (\ref{4.17}). Besides that, it involves arbitrary coupling constants $C^\lambda$ and $\bar C^{-\lambda}$, entering via 
the leading-order charges $q_1[P^-]$ and $q_1[J^{x-}]$. Accordingly, (\ref{4.25x1}) will be regarded as a linear differential equation on $A^{\lambda_1\lambda_2}$ and $a^{\lambda_1\lambda_2}$ with lower-order contributions viewed as sources. 
The constraint associated with $[P^-,J^{\bar{x}-}]=0$ will be treated analogously. It should be  kept in mind, that the unknown functions 
$A^{\lambda_1\lambda_2}$, $a^{\lambda_1\lambda_2}$ and $\bar a^{\lambda_1\lambda_2}$ also satisfy the kinematic constraints, which are given by  (\ref{21apr1x1}) with $n=2$.

The second-order consistency conditions involve fields quadratically. Below we will rename the coordinates of these two fields as $z_1^{\perp}=x^{\perp}$, $z_2^{\perp}=y^{\perp}$.
In these terms the first dynamical consistency condition reads
 \begin{equation}
  	\label{20feb5}
 \begin{split}
 	&a^{\lambda_1 \lambda_2}\left[\left(\frac{m^2}{2}+\sigma_{x}\sigma_{\bar x}\right)s_x+\left(\frac{m^2}{2}+\sigma_{y}\sigma_{\bar y}\right)s_y\right]\\
 	&\qquad+\left(s_x\sigma_{\bar x}\frac{\partial }{\partial s_x}
	+s_y\sigma_{\bar y}\frac{\partial }{\partial s_y}
 	-\sigma_{\bar x}^2 \frac{\partial}{\partial \sigma_{ \bar x}}
	-\sigma_{\bar y}^2 \frac{\partial}{\partial \sigma_{ \bar y}}\right.\\
	&\qquad\qquad\qquad  \left.
 	-\frac{m^2}{2}\frac{\partial}{\partial \sigma_{x}}
	-\frac{m^2}{2}\frac{\partial}{\partial \sigma_{y}}
	-\l_1\sigma_{\bar x}-\l_2\sigma_{\bar y}
 	\right)A^{\lambda_1 \lambda_2}
	=S^{\lambda_1\lambda_2}.
	\end{split}
 \end{equation}
The source term has the form 
\begin{equation}
		S^{\l_1\l_2}=S_1^{\l_1\l_2}+S_2^{\l_1\l_2}+S_3^{\l_1\l_2}+S_4^{\l_1\l_2}, 
	\end{equation}
	where
	\begin{equation}
		\begin{split}
			&S_1^{\l_1\l_2}=
			\frac{1}{2}C^{\lambda_1}C^{\lambda_2}\sigma_{x}^{\lambda_1}\sigma_{ y}^{\lambda_2-2}\lambda_2(\lambda_2-1)\frac{s_x(\sigma_{ x}-\sigma_y)}{ s_y}\\
			&\qquad\qquad \qquad\qquad \qquad\qquad+\frac{1}{2}C^{\lambda_1}C^{\lambda_2}\sigma_{y}^{\lambda_2}\sigma_{ x}^{\lambda_1-2}\lambda_1(\lambda_1-1)\frac{s_y(\sigma_{ y}-\sigma_x)}{ s_x},\\
			&S_2^{\l_1\l_2}=\bar C^{-\lambda_1} C^{\lambda_2}
			\sigma_{\bar x}^{-\lambda_1-1}\sigma_y^{\lambda_2-2}\Big[\lambda_1\lambda_2\sigma_{y}(\sigma_{\bar y}-\sigma_{\bar x})-\sigma_{\bar x}\frac{\lambda_2(\lambda_2-1)s_x}{s_y}(\sigma_{ y}-\sigma_x)\Big],
			\\
			&S_3^{\l_1\l_2}=\sum \limits_{\l}(-1)^{\l-1}\frac{3}{2} C^{\lambda} \ C^{-\lambda\lambda_1\lambda_2}\frac{
				(s_x\sigma_x+s_y \sigma_y)^{\l-1} (\sigma_y-\sigma_x)
				^{-\l+\l_1+\l_2-1}}
			{s^{\lambda-\lambda_2+1}_x s^{\l-\l_1+1}_y}\\
			&\qquad\qquad\qquad\qquad\qquad\qquad \qquad \Big[s_x\sigma_x\ (\l-\lambda_1+\lambda_2)+s_y \sigma_y (-\l-\lambda_1+\lambda_2)\Big],\\
			&S_4^{\l_1\l_2}=\sum \limits_{\lambda}(-1)^{\l+1}\frac{3}{2} \bar C^{-\l} \ C^{-\lambda\lambda_1\lambda_2}\frac{(\sigma_{\bar x} s_x+\sigma_{\bar y} s_y)^{-\l} (\sigma_y-\sigma_x)	^{-\lambda+\lambda_1+\lambda_2-1}}{s^{\lambda-\lambda_2+1}_x s^{\lambda-\lambda_1+1}_y(s_x+s_y)^{-2\l+1}}\\
			&\qquad\qquad\qquad\qquad\qquad\qquad\qquad \qquad\Big[s_x\ (\lambda-\lambda_1+\lambda_2)+s_y  (-\lambda-\lambda_1+\lambda_2)\Big],	
		\end{split}
	\end{equation}
	
	Analogously, the second dynamical consistency condition is 
	 \begin{equation}
  	\label{22feb3}
 \begin{split}
 	&\bar{a}^{\lambda_1 \lambda_2}\left[\left(\frac{m^2}{2}+\sigma_{x}\sigma_{\bar x}\right)s_x+\left(\frac{m^2}{2}+\sigma_{y}\sigma_{\bar y}\right)s_y\right]\\\
 	&\qquad+\left(s_x\sigma_{ x}\frac{\partial }{\partial s_x}
	+s_y\sigma_{ y}\frac{\partial }{\partial s_y}
 	-\sigma_{ x}^2 \frac{\partial}{\partial \sigma_{  x}}
	-\sigma_{y}^2 \frac{\partial}{\partial \sigma_{  y}}\right.\\
	&\qquad\qquad\qquad \left.
 	-\frac{m^2}{2}\frac{\partial}{\partial \sigma_{\bar x}}
	-\frac{m^2}{2}\frac{\partial}{\partial \sigma_{ \bar y}}
	+\l_1\sigma_{ x}+\l_2\sigma_{ y}
 	\right)A^{\lambda_1 \lambda_2}
	=\bar{S}^{\lambda_1\lambda_2}
	\end{split}
 \end{equation}
with the source	
\begin{equation}
	\bar	S^{\l_1\l_2}=\bar S_1^{\l_1\l_2}+\bar S_2^{\l_1\l_2}+\bar S_3^{\l_1\l_2},
\end{equation}
where
\begin{equation}
	\begin{split}
		\bar S_1^{\l_1\l_2}&=-\frac{1}{2}\bar C^{-\lambda_1}\bar C^{-\lambda_2}\sigma_{\bar x}^{-\lambda_1}\sigma_{\bar  y}^{-\lambda_2-2}\lambda_2(\lambda_2+1)\frac{s_x(\sigma_{\bar  x}-\sigma_{\bar y})}{ s_y}\\
		&	\qquad\qquad\qquad\qquad\qquad\qquad-\frac{1}{2}\bar C^{-\lambda_1}\bar C^{-\lambda_2}\sigma_{\bar x}^{-\lambda_2}\sigma_{\bar  y}^{-\lambda_1-2}\lambda_1(\lambda_1+1)\frac{s_x(\sigma_{\bar  y}-\sigma_{\bar x})}{ s_y},\\
		\bar S_2^{\l_1\l_2}&=C^{\lambda_1} \bar C^{-\lambda_2}
		\sigma_{ x}^{\lambda_1-1}\sigma_{\bar y }^{-\lambda_2-2}\Big[\lambda_1\lambda_2\sigma_{\bar y}(\sigma_{ x}-\sigma_{ y})+\sigma_{  x}\frac{\lambda_2(\lambda_2+1)s_x}{s_y}(\sigma_{\bar  x}-\sigma_{\bar y})\Big],\\
		\bar S_3^{\l_1\l_2}&=\sum \limits_{\l}\l (-1)^{\l-1}3 \bar C^{-\l} \ C^{-\lambda\lambda_1\lambda_2}\frac{
			(s_x\sigma_{\bar x}+s_y \sigma_{\bar y})^{-\l-1}(s_x+s_y)^{2\l-1} (\sigma_y-\sigma_x)
			^{-\l+\l_1+\l_2}}
		{s^{\l-\l_2}_x s^{\l-\l_1}_y}.
	\end{split}
\end{equation}	

Finally, the homogeneity constraints imposed by the kinematical consistency conditions are
\begin{equation}
\label{4.5555}
\begin{split}
\left(\sigma_{x}\frac{\partial}{\partial \sigma_{x}}-\sigma_{\bar x}\frac{\partial}{\partial \sigma_{\bar x}}+\sigma_{y}\frac{\partial}{\partial \sigma_{y}}-\sigma_{\bar y}\frac{\partial}{\partial \sigma_{\bar y}}-\l_1-\l_2 \right)A^{\l_1\l_2}=0,\\
\left(\sigma_{x}\frac{\partial}{\partial \sigma_{x}}-\sigma_{\bar x}\frac{\partial}{\partial \sigma_{\bar x}}+\sigma_{y}\frac{\partial}{\partial \sigma_{y}}-\sigma_{\bar y}\frac{\partial}{\partial \sigma_{\bar y}}-\l_1-\l_2 +1\right)a^{\l_1\l_2}=0,\\
\left(\sigma_{x}\frac{\partial}{\partial \sigma_{x}}-\sigma_{\bar x}\frac{\partial}{\partial \sigma_{\bar x}}+\sigma_{y}\frac{\partial}{\partial \sigma_{y}}-\sigma_{\bar y}\frac{\partial}{\partial \sigma_{\bar y}}-\l_1-\l_2-1 \right)\bar{a}^{\l_1\l_2}=0.
\end{split}
\end{equation}

\section{General solution on-shell}
\label{sec:4}

In the present section we will give the general solution to the consistency  conditions (\ref{20feb5}), (\ref{22feb3}) and 
(\ref{4.5555}) on-shell.

We start by noting that 
\begin{equation}
\begin{split}
			&\left(s_x\sigma_{\bar x}\frac{\partial }{\partial s_x}+s_y\sigma_{\bar y}\frac{\partial }{\partial s_y}
		-\sigma_{\bar x}^2 \frac{\partial}{\partial \sigma_{ \bar x}}
		-\sigma_{\bar y}^2 \frac{\partial}{\partial \sigma_{ \bar y}}-\frac{m^2}{2}\left(\frac{\partial}{\partial \sigma_{x}}+\frac{\partial}{\partial \sigma_{y}}\right)
		\right)k=0,\\
		&\left(s_x\sigma_{ x}\frac{\partial }{\partial s_x}+s_y\sigma_{ y}\frac{\partial }{\partial s_y}
		-\sigma_{ x}^2 \frac{\partial}{\partial \sigma_{  x}}
		-\sigma_{ y}^2 \frac{\partial}{\partial \sigma_{  y}}-\frac{m^2}{2}\left(\frac{\partial}{\partial \sigma_{\bar x}}+\frac{\partial}{\partial \sigma_{\bar y}}\right)
		\right)k=0,\\
		&\left(\sigma_{x}\frac{\partial}{\partial \sigma_{x}}-\sigma_{\bar x}\frac{\partial}{\partial \sigma_{\bar x}}+\sigma_{y}\frac{\partial}{\partial \sigma_{y}}-\sigma_{\bar y}\frac{\partial}{\partial \sigma_{\bar y}}\right)k=0,
	\end{split}
	\end{equation}
where 
	\begin{equation}
	\label{29apr1}
		k\equiv \Big(\frac{m^2}{2}(s_x+s_y)+s_x\sigma_{x}\sigma_{\bar x}+s_y\sigma_{y}\sigma_{\bar y}\Big).
	\end{equation}
In other words,  the differential operators occurring in (\ref{20feb5}), (\ref{22feb3}) and 
(\ref{4.5555}) only involve derivatives in the directions, which are tangential  to hypersurfaces with constant $k$, thus, for different $k$ these equations can be considered separately. 

We then focus on (\ref{20feb5}), (\ref{22feb3}) and 
(\ref{4.5555}) for
\begin{equation}
\label{21apr1}
k=0.
\end{equation}
It will be shown below, that once energy conservation is taken into account (\ref{21apr1}) implies that the charges are evaluated on-shell. 

Functions $a$ and $\bar a$ enter into the consistency conditions in combinations $k \cdot a$ and $k\cdot \bar a$. These combinations are vanishing on $k=0$ unless $a$ and $\bar a$ have $1/k$-type poles. The latter is inconsistent with locality, thus, by setting $k=0$ we remove $a$ and $\bar a$ from the consistency conditions. As a result, we find  
\begin{equation}
	\label{4.110}
\begin{split}
		&\Big(s_x\sigma_{\bar x}\frac{\partial }{\partial s_x}+s_y\sigma_{\bar y}\frac{\partial }{\partial s_y}
		-\sigma_{\bar x}^2 \frac{\partial}{\partial \sigma_{ \bar x}}
		-\sigma_{\bar y}^2 \frac{\partial}{\partial \sigma_{ \bar y}}\\
		&\qquad\qquad\qquad\qquad-\frac{m^2}{2}\left(\frac{\partial}{\partial \sigma_{x}}+\frac{\partial}{\partial \sigma_{y}}\right)-\l_1\sigma_{\bar x}-\l_2\sigma_{\bar y}
		\Big)A^{\lambda_1\lambda_2}|_{k=0}=S^{\lambda_1\lambda_2}|_{k=0},	
		\\
		&\Big(s_x\sigma_{ x}\frac{\partial }{\partial s_x}+s_y\sigma_{ y}\frac{\partial }{\partial s_y}
		-\sigma_{ x}^2 \frac{\partial}{\partial \sigma_{  x}}
		-\sigma_{ y}^2 \frac{\partial}{\partial \sigma_{  y}}\\
		&\qquad\qquad\qquad\qquad-\frac{m^2}{2}\left(\frac{\partial}{\partial \sigma_{\bar x}}+\frac{\partial}{\partial \sigma_{\bar y}}\right)+\l_1\sigma_{ x}+\l_2\sigma_{ y}
		\Big)A^{\lambda_1\lambda_2}|_{k=0}=\bar S^{\lambda_1\lambda_2}|_{k=0},
		\\
		&\left(\sigma_{x}\frac{\partial}{\partial \sigma_{x}}-\sigma_{\bar x}\frac{\partial}{\partial \sigma_{\bar x}}+\sigma_{y}\frac{\partial}{\partial \sigma_{y}}-\sigma_{\bar y}\frac{\partial}{\partial \sigma_{\bar y}}-\l_1-\l_2 \right)A^{\l_1\l_2}|_{k=0}=0.
	\end{split}
	\end{equation}
	
	Equations (\ref{4.110}) present a system of non-homogeneous differential equations for $A$ on $k=0$. Accordingly, its solution has the form of the sum of the general solution of the homogeneous equations and  a particular solution of the non-homogeneous equations. 

The general solution of the homogeneous equations can be found straightforwardly by consecutively solving each equation and plugging the solution into the remaining ones. As a result, we find
\begin{equation}
\label{21apr2}
A_h^{\lambda_1\lambda_2}=(l^{++})^{\frac{\lambda_1+\lambda_2}{2}} (l^{+-})^{\frac{\lambda_1-\lambda_2}{2}}   f^{\lambda_1\lambda_2}(k_1,k_3),
\end{equation}
where 
\begin{equation}
	\label{kkkk}
\begin{split}
		k_1\equiv s_x\left(\sigma_{  x}\sigma_{ \bar x}+\frac{m^2}{2}\right), \qquad
		k_3\equiv s_xs_y(\sigma_{  x}-\sigma_{  y})(\sigma_{ \bar  x}-\sigma_{ \bar  y}),\\
		l^{++} \equiv  s_x s_y(\sigma_x-\sigma_y)^2, \qquad l^{+-} \equiv s_xs_y\left(\sigma_{  x}\sigma_{ \bar y}+\frac{m^2}{2}\right)^2
		\end{split}
	\end{equation}
and $f$ is an arbitrary function. Along with $k_1$, it is natural to introduce
\begin{equation}
\label{4june1}
k_2\equiv s_y\left(\sigma_{  y}\sigma_{ \bar y}+\frac{m^2}{2}\right),
\end{equation}
which satisfies
\begin{equation}
\label{4jun2}
k=k_1+k_2,
\end{equation}
thus, on-shell (\ref{21apr1}), $k_1$ in (\ref{21apr2}) can be replaced with $-k_2$. In fact, if $f$  depends on both $k_1$ and $k_2$,  (\ref{21apr2}) gives the general solution of the homogeneous equations off-shell.

From (\ref{21apr2}) it is not hard to see that each $l^{++}$ factor is responsible for raising both helicities by one. At the same time, 
$l^{+-}$ raises $\lambda_1$ by one and lowers $\lambda_2$ by one. Note that along with $l^{++}$ and $l^{+-}$, we can also consider 
\begin{equation}
\label{21apr3}
l^{--} \equiv  s_x s_y(\sigma_{\bar x}-\sigma_{\bar y})^2, \qquad l^{-+} \equiv s_xs_y\left(\sigma_{  \bar{x}}\sigma_{  y}+\frac{m^2}{2}\right)^2,
\end{equation}
where the first factor lowers both helicities by one, while the second one lowers the first and raises the second helicity by one respectively. Considering that 
$l^{++} l^{--}$ and $l^{+-} l^{-+}$ keep helicities intact, consistency with (\ref{21apr2}) implies that these are both expressible in terms of $k_1$, $k_2$ and $k_3$ off-shell and in terms of $k_1$ and $k_3$ on-shell. This is, indeed, the case, namely, 
\begin{equation}
\label{21apr4}
l^{++} l^{--} =k_3^2, \qquad 
l^{+-} l^{-+}=k_1k_2-\frac{m^2}{2}k_3.
\end{equation} 
To keep locality of $A_h$ in (\ref{21apr2}) manifest, one should avoid negative powers of $l^{++}$ and $l^{+-}$. Depending on the values of $\lambda_1$ and $\lambda_2$ achieving that may require trading $l^{++}$ for $(l^{--})^{-1}$ and $l^{+-}$ for $(l^{-+})^{-1}$.

Considering that we are dealing with linear differential equations, a particular solution of the non-homogeneous equations can also be obtained in a straightforward manner by solving one equation after another. In the following we will study whether the light-cone consistency conditions have solutions which are free of singularities. To do that, we will start from a particular solution of the non-homogeneous equations and explore whether its singularities can be removed by adding solutions of the homogeneous equations. For this procedure to work efficiently, it is important to start with a simple particular solution of non-homogeneous equations, which has as few singularities as possible. 
The direct method of solving the non-homogeneous equations does not necessarily lead to simple particular solutions. Moreover, since we are dealing with a system of equations, an unfortunate choice of a particular solution when solving equations at the first steps of the procedure, may considerably complicate the options for particular solutions at the following steps. 
Due to that, when looking for a particular solution of (\ref{4.110}) it is beneficial to have some additional considerations, which would help guessing it in a convenient form. 

Relying on the anticipated connection of the light-cone approach with scattering amplitudes, 
which will be further discussed below, we found the following particular  solution of   (\ref{4.110})
\begin{equation}
		A_{nh}^{\l_1\l_2}=A_{pp}^{\l_1\l_2}+A_{nf}^{\l_1\l_2}+A_{pf}^{\l_1\l_2},
		\label{4.18}
	\end{equation}
	where 
\begin{eqnarray}
		&&A^{\l_1\l_2}_{pf}=\frac{3}{2}\sum_{\lambda} C^{\l_1\l_2-\l}C^{\l}(s_x\sigma_x+s_y\sigma_y)^{\l}\frac{(\sigma_x-\sigma_y)^{\l_1+\l_2-\l-1}}{(\sigma_{\bar x}-\sigma_{\bar y})s_x^{\l-\l_2+1} s_y^{\l-\l_1+1}},\label{4.2}\\
		&&A^{\l_1\l_2}_{nf}=-\frac{3}{2}\sum_{\lambda}  C^{\l_1\l_2-\l}\bar C^{-\l}\frac{(\sigma_x-\sigma_y)^{\l_1+\l_2-\l-1}(s_x+s_y)^{2\l}}{(s_x\sigma_{\bar x}+s_y\sigma_{\bar y})^{\l}(\sigma_{\bar x}-\sigma_{\bar y})s_x^{\l-\l_2+1} s_y^{\l-\l_1+1}},\label{4.22}\\
		&&A^{\l_1\l_2}_{pp}=\frac{1}{2}C^{\l_1}C^{\l_2}\sigma_x^{\l_1}\sigma_y^{\l_2}\left(\frac{\sigma_{ x}-\sigma_{ y}}{\sigma_x\sigma_y}
		\frac{(\l_2s_x\sigma_x+\l_1s_y\sigma_y)}{k_1}-\frac{k_3}{k_1^2}\right).
		\label{4.21}
	\end{eqnarray}
As will be explained later, each contribution in (\ref{4.18}) can be related to an exchange of a different type. 
Namely, $A^{\l_1\l_2}_{pp}$ can be associated with the exchange diagram in which the point particle is being exchanged, 
while $A^{\l_1\l_2}_{pf}$ and $A^{\l_1\l_2}_{nf}$ can be regarded as contributions from positive- and negative-helicity field exchanges.
We would like to remark that $A^{\l_1\l_2}_{pp}$ gives the necessary particular solution only for 
$\l_1\geq0$ or $\l_2\geq0$, which will be enough for our analysis in what follows. Finally, let us remark that (\ref{4.18}) provides a solution to (\ref{20feb5}), (\ref{22feb3}) and 
(\ref{4.5555}) only up to terms that vanish on $k=0$.

\subsection{On-shell condition}

Before proceeding to the next section, in which we will study whether the general solution presented above gives any local vertices, let us clarify the meaning of the condition 
(\ref{21apr1}). Namely, we will show that once the energy conservation or, equivalently, the freedom of integration by parts is taken into account, (\ref{21apr1}) implies that the charges we are dealing with are evaluated on-shell in the sense that the free equations of motion are taken into account.

In order to do that, we consider a point particle vertex associated with the second-order correction to the Hamiltonian, see (\ref{4.17})
\begin{equation}
\label{29apr2}
\begin{split}
S_{pp|2} = -\int dx^+ \sum_{\lambda_1 \lambda_2} \frac{1}{p^+}A^{\lambda_1\lambda_2}(\sigma_x,\sigma_y,\sigma_{\bar x}, \sigma_{\bar y},s_x,s_y)\Phi^{\lambda_1}(x^\perp)\Phi^{\lambda_2}(y^\perp)\Big|_{x^\perp=y^\perp}.
\end{split}
\end{equation}
As in field theory, the point particle Lagrangian is defined up to total-derivative terms 
\begin{equation}
\label{29apr3}
\begin{split}
\delta_{t.d.}L_{pp|2} = - \sum_{\lambda_1,\lambda_2} 
\frac{d}{dx^+ } \big(\beta^{\lambda_1\lambda_2}(p^\perp,\partial_x^\perp,\partial_y^\perp)\Phi^{\lambda_1}(x^\perp)\Phi^{\lambda_2}(y^\perp) \big)\Big|_{x^\perp=y^\perp}
\end{split}
\end{equation}
with any $\beta$, which is consistent with the kinematical constraints. 
The total time derivative of the field is 
\begin{equation}
\label{7jun10}
	\frac{d\Phi^{\lambda_1}}{dx^+}=\frac{\partial \Phi^{\lambda_1}}{\partial x^+}  + \frac{\partial \Phi^{\lambda_1}}{\partial x}\dot x  + \frac{\partial \Phi^{\lambda_1}}{\partial \bar x}\dot{\bar x}+\frac{\partial \Phi^{\lambda_1}}{\partial x^-}\dot x^-.
\end{equation}
By employing the free field equations of motion to eliminate $\partial_+\Phi$ as well as eliminating point particle velocities in terms of momenta via its free equations of motion, we find
\begin{equation}
	\frac{d\Phi^{\lambda_1}}{d x^+}\approx -\frac{\partial^+_x}{(p^+)^2} \left(\sigma_x \sigma_{\bar x} +\frac{m^2}{2}\right)\Phi^{\lambda_1}
	=-\frac{k_1}{p^+}\Phi^{\lambda_1},
	\label{4.36}
	\end{equation}
where "$\approx$" highlights that the free equations of motion were used. Besides that, 
\begin{equation}
	\frac{d\beta}{d x^+}\approx 0
	\label{29apr4}
	\end{equation}
due to the fact that $\beta$ is $x^\perp$-independent -- as a consequence of the kinematical constraints -- as well as $\dot{p}^\perp\approx 0$. 
Substituting (\ref{4.36}) and (\ref{29apr4}) into (\ref{29apr3}), we find that 
\begin{equation}
\label{30apr5}
\delta_{t.d.}L_{pp|2} \approx \sum_{\lambda_1,\lambda_2} 
\frac{\beta^{\lambda_1\lambda_2}}{p^+}
(k_1+k_2)
\Phi^{\lambda_1}(x^\perp)\Phi^{\lambda_2}(y^\perp) \Big|_{x^\perp=y^\perp}.
\end{equation}
By comparing this with (\ref{29apr2}), we find that
\begin{equation}
\label{29apr6}
A^{\lambda_1\lambda_2}
\approx A^{\lambda_1\lambda_2} -k\beta^{\lambda_1\lambda_2}.
\end{equation}

Thus, in terms of $A$ on-shell trivial terms manifest themselves as terms proportional to $k$. Equivalently, point particle vertex (\ref{29apr2}) is put on-shell once (\ref{21apr1})
is imposed, as we intended to show\footnote{Note, that a condition analogous to  (\ref{21apr1}) was referred to as energy conservation in \cite{Metsaev:1991nb}. As we demonstrated, both energy conservation and the on-shell conditions should be taken into account in order to derive (\ref{21apr1}). Out of the two conditions, when referring to (\ref{21apr1}) we would like to emphasise its on-shell nature, because it is an additional condition that one imposes when passing to amplitudes, while energy conservation and the freedom of integration by parts are built in already at the action level.}. 
It is well-known that for field theories on-shell trivial vertices give fake interactions in the sense that these can be removed by field redefinition. In  \cite{Ivanovskiy:2023aay} it was shown that an analogous conclusion also applies to worldline theories. In other words, by imposing  (\ref{21apr1}), we remove fake interaction vertices of point particles, which are quadratic in fields.

\section{Removing singularities}
\label{sec:5}

In the previous section we found the general solution to the light-cone consistency conditions (\ref{20feb5}), (\ref{22feb3}) and 
(\ref{4.5555}). It has the form of the sum of the general solution of the homogeneous equations (\ref{21apr2}) and  the particular solution of the non-homogeneous equations (\ref{4.18}), 
the latter consisting of the three contributions explicitly given in (\ref{4.2})-(\ref{4.21})
\begin{equation}
\label{24apr1}
A^{\lambda_1\lambda_2} = A_{h}^{\lambda_1\lambda_2} +A_{nh}^{\lambda_1\lambda_2} = 
 A_{h}^{\lambda_1\lambda_2} +A^{\l_1\l_2}_{pf}+ A^{\l_1\l_2}_{nf}+A^{\l_1\l_2}_{pp}.
\end{equation}
As one can see, the particular solutions (\ref{4.2}) and (\ref{4.22}) have poles at $\sigma_{\bar x}=\sigma_{\bar y}$, while the particular solution (\ref{4.21}) has a second-order pole at
$\sigma_{x}\sigma_{\bar x}=-\frac{m^2}{2}$. Locality requires that the total charge defined by (\ref{24apr1}) is free of singularities in $\sigma_x$, $\sigma_{y}$, $\sigma_{\bar x}$ and $\sigma_{\bar y}$. This means that for a solution to be local, singularities contributed by the non-homogeneous terms should be cancelled by a properly chosen homogeneous solution.
Below we will look for local solutions in (\ref{24apr1}). Before going to concrete computations, let us explain how this analysis works at the qualitative level.

Firstly, it should be noted that the general solution of the homogeneous equations (\ref{21apr2}) has a rather special form: for a fixed set of helicities it effectively depends only on two variables $k_1$ and $k_3$. Accordingly, (\ref{24apr1}) can give local solutions only if singular parts of $A_{nh}^{\lambda_1\lambda_2}$ are expressible in the form (\ref{21apr2}). 
In particular, $A^{\l_1\l_2}_{pf}+ A^{\l_1\l_2}_{nf}$ at  $\sigma_{\bar x}\to\sigma_{\bar y}$ should be expressible as
\begin{equation}
\label{24apr2}
A^{\l_1\l_2}_{pf}+ A^{\l_1\l_2}_{nf} = (l^{++})^{\frac{\lambda_1+\lambda_2}{2}} (l^{+-})^{\frac{\lambda_1-\lambda_2}{2}} \left(\frac{r_{3}(k_1)}{k_3}+ O\left((\sigma_{\bar x}-\sigma_{\bar y})^0\right) \right).
\end{equation}
 In other words, once the helicity factors are removed, not only the denominators of
$A^{\l_1\l_2}_{pf}+ A^{\l_1\l_2}_{nf}$
should feature $(\sigma_{\bar x}-\sigma_{\bar y})$  together with other factors present in $k_3$, see (\ref{kkkk}), the residue at this pole should be also expressible as a function of $k_1$ alone.
 The same refers to $A^{\l_1\l_2}_{pp}$: local solutions may exist only if its singular part at $\sigma_{x}\sigma_{\bar x}\to -\frac{m^2}{2}$ has the form 
\begin{equation}
\label{24apr3}
A^{\l_1\l_2}_{pp} = (l^{++})^{\frac{\lambda_1+\lambda_2}{2}} (l^{+-})^{\frac{\lambda_1-\lambda_2}{2}} \left(\frac{r_{1}(k_3)}{k_1^2}+\frac{r'_{1}(k_3)}{k_1}+ O\left((\sigma_{\bar x}\sigma_{\bar x}+\frac{m^2}{2})^0\right) \right).
\end{equation}

As we will see below, conditions (\ref{24apr2}) and (\ref{24apr3}) do not result in any constraints in the sense that these are satisfied automatically for all  particular solutions (\ref{4.2})-(\ref{4.21}). This feature has a simple explanation in the language of amplitudes. Namely, it is connected to the fact that in contrast to exchanges, which do not give consistent amplitudes -- depending on the approach, these are either not Lorentz invariant or not gauge invariant -- their singular parts are, as a consequence of unitarity, which requires that residues of exchanges factorise into consistent lower-point amplitudes.

Once  (\ref{24apr2}) and (\ref{24apr3}) are verified, one encounters the second consistency condition imposed by locality, which turns out to be non-trivial. 
Namely, (\ref{24apr2}) implies, that the homogeneous solution (\ref{21apr2}), that needs to be added to achieve local $A^{\lambda_1\lambda_2}$ in (\ref{24apr1})
behaves as
\begin{equation}
\label{24apr4}
f(k_1,k_3) = -\frac{r_{3}(k_1)}{k_3}+ O\left(k_3^0\right), \qquad k_3 \to 0.
\end{equation}
At the same time, (\ref{24apr3}) implies, that for local solutions one should also require
\begin{equation}
\label{24apr5}
f(k_1,k_3) = -\frac{r_{1}(k_3)}{k_1^2}-\frac{r'_{1}(k_3)}{k_1}+ O\left(k_1^0\right), \qquad k_1 \to 0.
\end{equation}
Below we will find that $r'_{1}(k_3)=0$. With this in mind, consistency of (\ref{24apr4}) with  (\ref{24apr5}) in the limit in which both $k_1$ and $k_3$ go to zero implies that $r_1(k_3)$ and $r_3(k_1)$ have the following singular parts
\begin{equation}
\label{24apr6}
r_1(k_3) = \frac{\alpha}{k_3}+O(k_3^0), \qquad r_3(k_1) = \frac{\alpha}{k_1^2}+O(k_1^0).
\end{equation}
Note that (\ref{24apr6}) not only requires that coefficients $\alpha$ occurring in both expansions are the same, it also implies that other singular terms in each expansion are absent. Equation (\ref{24apr6}) with $r_1$ and $r_3$ expressed in terms of lower-order coupling constants $C$ and $\bar C$ gives the  consistency condition on these coupling constants, which needs to be satisfied, otherwise, local interactions between a point particle and massless fields do not exist. 

\subsection{Self-dual gravity}

To test our approach, we start by considering a well-understood case of self-dual gravity. To this end, we focus on the solution (\ref{24apr1}) in the sector $\lambda_1=\lambda_2 =2$. It has the form
\begin{equation}
\label{24apr7}
\begin{split}
&A^{2,2} = (l^{++})^2 \left[ f^{2,2}(k_1,k_3)+\frac{3}{2} l C^2\left(\frac{s_x\sigma_x + s_y \sigma_y}{s_x s_y (\sigma_{ x} - \sigma_{ y})}\right)^2
 \frac{1}{k_3} \right.
 \\
&\qquad\qquad  \qquad  \left.+
 \frac{1}{2}(C^2)^2
 \left(\frac{\sigma_x\sigma_y}{s_xs_y (\sigma_x-\sigma_y)^2} \right)^2 \left(2\frac{\sigma_x-\sigma_y}{\sigma_x \sigma_y}  \frac{s_x \sigma_x+s_y \sigma_y}{k_1} - \frac{k_3}{(k_1)^2} \right)
  \right].
  \end{split}
\end{equation}

We first consider the limit $\sigma_{\bar x} \to \sigma_{\bar y}$. In this limit $k_3 \to 0$, so the second term in the squared brackets is singular, while the third one is finite. 
To evaluate the residue of the second term we use that
\begin{equation}
\label{24apr8}
\frac{s_x\sigma_x + s_y \sigma_y}{s_x s_y (\sigma_{ x} - \sigma_{ y})} = \frac{\frac{m^2}{2}}{\frac{s_x s_y(\sigma_{\bar x}-\sigma_{\bar y})\sigma_x\sigma_y}{s_x \sigma_x +s_y\sigma_y} - k_1},
\end{equation}
which can be easily verified by a direct computation. Accordingly, 
\begin{equation}
\label{24apr9}
\frac{3}{2}\ l C^2\left(\frac{s_x\sigma_x + s_y \sigma_y}{s_x s_y (\sigma_{ x} - \sigma_{ y})}\right)^2
 \frac{1}{k_3} =
 \frac{3}{2}\ l C^2 \left(\frac{m^2}{2} \right)^2 \frac{1}{k_1^2}\frac{1}{k_3} + O(k_3^0), \qquad k_3\to 0.
\end{equation}
To remove this singularity, the homogeneous solution should behave as
\begin{equation}
\label{24apr10}
f^{2,2}(k_1,k_3)= -  l C^2 \left(\frac{m^2}{2} \right)^2 \frac{3}{2k_1^2 k_3} + O(k_3^0), \qquad k_3\to 0.
\end{equation}

Next, we consider the limit $\sigma_{x}\sigma_{\bar x}\to -\frac{m^2}{2}$. In this limit $k_1 \to 0$, so the third term in the squared brackets in (\ref{24apr7}) is singular, while the second one is regular. To evaluate the singular part of the third term, we use (\ref{24apr8}) together with
\begin{equation}
\label{23apr9}
\frac{\sigma_x\sigma_y}{s_xs_y (\sigma_x-\sigma_y)^2} = \frac{m^2/2}{k_3-k_1 \frac{(\sigma_x -\sigma_y)(s_x \sigma_x +s_y \sigma_y)}{\sigma_x \sigma_y}},
\end{equation}
which again can be easily checked. Accordingly, we find
\begin{equation}
\label{24apr11}
\begin{split}
& \frac{1}{2}(C^2)^2
 \left(\frac{\sigma_x\sigma_y}{s_xs_y (\sigma_x-\sigma_y)^2} \right)^2 \left(2\frac{\sigma_x-\sigma_y}{\sigma_x \sigma_y}  \frac{s_x \sigma_x+s_y \sigma_y}{k_1} - \frac{k_3}{(k_1)^2} \right)\\
&\qquad\qquad\qquad\qquad \qquad\qquad    =-
 (C^2)^2\left(\frac{m^2}{2} \right)^2 \frac{1}{2k_1^2 k_3 } + O(k_1^0), \qquad k_1\to 0.
 \end{split}
\end{equation}
Note, that the first order pole in $k_1$ is absent. To cancel this singularity, the homogeneous solution should satisfy 
\begin{equation}
\label{24apr12}
f^{2,2}(k_1,k_3)=  (C^2)^2 \left(\frac{m^2}{2} \right)^2 \frac{1}{2k_1^2 k_3} + O(k_3^0), \qquad k_3\to 0.
\end{equation}

Consistency of (\ref{24apr10}) and (\ref{24apr12}) in the limit in which both $k_1$ and $k_3$ go to zero requires
\begin{equation}
\label{24apr13}
(C^2)^2 = -3 l C^2.
\end{equation}
In other words, non-trivial local interactions of a point particle with self-dual gravity are possible and the associated coupling constant $C^2$ is related to the coupling constant on the field theory side $l$ -- which is, in turn, related to  Newton's coupling constant -- by 
\begin{equation}
\label{24apr14}
C^2 = - 3l.
\end{equation}
 This result is consistent with our findings for point particles in the massless case \cite{Ivanovskiy:2025kok}. 

\subsection{Chiral higher-spin theory}

Let us now consider the case of chiral higher-spin theory, for which the cubic coupling constants are given by (\ref{28feb4}). We will focus on the consistency conditions implied by locality in the sector with $\lambda_2 =0$ and $\lambda_1 \ge 0$. Then, the solution (\ref{24apr1}) reads
\begin{equation}
\label{24apr15}
\begin{split}
&A^{\lambda_1,0}=(l^{++})^{\frac{\lambda_1}{2}} (l^{+-})^{\frac{\lambda_1}{2}} \\
&\quad \left[ f^{\lambda_1,0}(k_1,k_3)+
\sum_{\lambda} \frac{3 l^{\lambda_1-\lambda -1}C^{\l}}{2\Gamma(\lambda_1 -\lambda)}
\left(\frac{s_x\sigma_x + s_y \sigma_y}{s_x s_y (\sigma_{ x} - \sigma_{ y})} \right)^\lambda \left(s_x(\sigma_x \sigma_{\bar y}+\frac{m^2}{2})\right)^{-\lambda_1} \frac{1}{k_3}
\right.\\
&\quad - \sum_{\lambda} \frac{3 l^{\lambda_1-\lambda -1}\bar{C}^{-\l}}{2\Gamma(\lambda_1 -\lambda)}
\left(\frac{(s_x+s_y)^2}{(s_x\sigma_{\bar x}+s_y\sigma_{\bar y})(\sigma_x-\sigma_y)s_x s_y} \right)^{\lambda}
 \left(s_x(\sigma_x \sigma_{\bar y}+\frac{m^2}{2})\right)^{-\lambda_1}\frac{1}{k_3}\\
&\qquad \qquad \;  \left.+ \frac{1}{2}C^{\lambda_1}C^0 
 \left(
\frac{\sigma_x}{s_x s_y (\sigma_x-\sigma_y)(\sigma_x\sigma_{\bar x} + \frac{m^2}{2})} \right)^{\lambda_1}
\left(\lambda_1\frac{s_y(\sigma_x-\sigma_y)}{\sigma_x}\frac{1}{k_1}-\frac{k_3}{k_1^2} \right)\right].
\end{split}
\end{equation}

We start by considering the limit $\sigma_{\bar x} \to \sigma_{\bar y}$, for which $k_3\to 0$. In this limit the second and the third terms are singular, while the last one is regular. To evaluate the singular terms in the given limit we use that
\begin{equation}
\label{24apr16}
\begin{split}
 s_x(\sigma_x \sigma_{\bar y}+\frac{m^2}{2}) &= k_1 + s_x\sigma_x (\sigma_{\bar y}-\sigma_{\bar x}),\\
 \frac{(s_x+s_y)^2}{(s_x\sigma_{\bar x}+s_y\sigma_{\bar y})(\sigma_x-\sigma_y)s_x s_y} &= 
\frac{1}{k_1 + (\sigma_{\bar y}-\sigma_{\bar x}) \frac{s_y s_x (s_y\sigma_x + s_x \sigma_y)}{(s_x+s_y)^2}}
 \end{split}
\end{equation}
together with (\ref{24apr8}). As a result, we find that the compensating homogeneous solution should behave as
\begin{equation}
\label{24apr17}
f^{\lambda_1,0}= - \sum_{\lambda} \frac{3 l^{\lambda_1-\lambda -1}C^{\l}}{2\Gamma(\lambda_1 -\lambda)} \left(-\frac{m^2}{2} \right)^\lambda \frac{1}{k_1^{\lambda+\lambda_1}}\frac{1}{k_3}+\sum_{\lambda} \frac{3 l^{\lambda_1-\lambda -1}\bar{C}^{-\l}}{2\Gamma(\lambda_1 -\lambda)} \frac{1}{k_1^{\lambda+\lambda_1}}\frac{1}{k_3}+O(k_3^0).
\end{equation}
Due to the presence of gamma functions the sum over $\lambda$ goes over $\lambda \le \lambda_1 -1$. Moreover, for chiral higher-spin theories without internal symmetries vertices with the total number of derivatives, which is $\lambda_1-\lambda$, odd are vanishing. Accordingly, the sum in (\ref{24apr17}) has non-vanishing contributions for $\lambda = \lambda_1-2-2n$ with $n$ being non-negative integer and the associated orders of poles in $k_1$ are $2\lambda_1-2n -2$.
Considering that the last term in (\ref{24apr15}) has only second order poles in $k_1$, consistency requires that the coupling constants $C^\lambda$ and $\bar{C}^{-\lambda}$
are  vanishing for $\lambda = 4-\lambda_1, 6-\lambda_1, \dots, \lambda_1 -2$. Since, $\lambda_1$ can be arbitrary, we conclude that
\begin{equation}
\label{25apr1}
C^{\lambda} = 0, \qquad \bar{C}^{-\lambda} =0, \qquad \forall  \lambda.
\end{equation}
Thus, local interactions of point particles with the chiral higher-spin theory do not exist. 

For completeness, let us consider the last term in (\ref{24apr15}) in the limit $\sigma_{x}\sigma_{\bar x}\to -\frac{m^2}{2}$.
One can show that
\begin{equation}
\label{25apr2}
\frac{\sigma_x}{s_x s_y (\sigma_x-\sigma_y)(\sigma_x\sigma_{\bar x} + \frac{m^2}{2})} = \frac{1}{\frac{s_y(\sigma_x-\sigma_y)}{\sigma_x} k_1 -k_3}.
\end{equation}
Accordingly, 
\begin{equation}
\label{25apr3}
\begin{split}
 & \frac{1}{2}C^{\lambda_1}C^0 \left(
\frac{\sigma_x}{s_x s_y (\sigma_x-\sigma_y)(\sigma_x\sigma_{\bar x} + \frac{m^2}{2})} \right)^{\lambda_1}
\left(\lambda_1\frac{s_y(\sigma_x-\sigma_y)}{\sigma_x}\frac{1}{k_1}-\frac{k_3}{k_1^2} \right)\\
& \qquad\qquad  \qquad\qquad  \qquad\qquad  \qquad\qquad   =(-1)^{\lambda_1-1} C^{\lambda_1}C^0 \frac{1}{2k_3^{\lambda_1-1}k_1^2}+O(k_1^0)
\end{split}
\end{equation}
and the compensating solution of homogeneous equations should behave as
\begin{equation}
\label{25apr3x1}
f^{\lambda_1,0}= - (-1)^{\lambda_1-1} C^{\lambda_1}C^0 \frac{1}{2k_3^{\lambda_1-1}k_1^2}+O(k_1^0).
\end{equation}

\subsection{Poisson chiral higher-spin theory}

Finally, we explore the coupling of the point particle to the Poisson chiral higher-spin theory. The field theory coupling constants are given by (\ref{28feb5}).
We start by considering the sector of consistency conditions with $\lambda_2=0$ and $\lambda_1\ge 0$. Then, solution (\ref{24apr1}) can be written as 
\begin{equation}
\label{25apr4}
\begin{split}
&A^{\lambda_1,0}=(l^{++})^{\frac{\lambda_1}{2}} (l^{+-})^{\frac{\lambda_1}{2}} \\
&\quad \left[ f^{\lambda_1,0}(k_1,k_3)+
{3 l C^{\l_1-2}}
\left(\frac{s_x\sigma_x + s_y \sigma_y}{s_x s_y (\sigma_{ x} - \sigma_{ y})} \right)^{\lambda_1-2} \left(s_x(\sigma_x \sigma_{\bar y}+\frac{m^2}{2})\right)^{-\lambda_1} \frac{1}{k_3}
\right.\\
&\quad -3 l\bar{C}^{-\l_1+2}
\left(\frac{(s_x+s_y)^2}{(s_x\sigma_{\bar x}+s_y\sigma_{\bar y})(\sigma_x-\sigma_y)s_x s_y} \right)^{\lambda_1-2}
 \left(s_x(\sigma_x \sigma_{\bar y}+\frac{m^2}{2})\right)^{-\lambda_1}\frac{1}{k_3}\\
&\qquad \quad  \left.+ \frac{1}{2} C^{\lambda_1}C^0 
 \left(
\frac{\sigma_x}{s_x s_y (\sigma_x-\sigma_y)(\sigma_x\sigma_{\bar x} + \frac{m^2}{2})} \right)^{\lambda_1}
\left(\lambda_1\frac{s_y(\sigma_x-\sigma_y)}{\sigma_x}\frac{1}{k_1}-\frac{k_3}{k_1^2} \right)\right].
\end{split}
\end{equation}

Again, in the limit $\sigma_{\bar x} \to \sigma_{\bar y}$  the second and the third terms in the brackets of (\ref{25apr4}) are singular. Proceeding as before, we find that to compensate this singularity in the total solution, one should have 
\begin{equation}
\label{25apr5}
f^{\lambda_1,0}= -  {3 lC^{\l_1-2}} \left(-\frac{m^2}{2} \right)^{\lambda_1-2} \frac{1}{k_1^{2\lambda_1-2}}\frac{1}{k_3}+ {3 l\bar{C}^{-\l_1+2}} \frac{1}{k_1^{2\lambda_1-2}}\frac{1}{k_3}+O(k_3^0).
\end{equation}
As for the limit $\sigma_{x}\sigma_{\bar x}\to -\frac{m^2}{2}$, it remains the same as in the chiral higher-spin theory case (\ref{25apr3x1}).

To avoid higher than the second order singularities in $k_1$  in (\ref{25apr5}) one should require
\begin{equation}
\label{26apr1}
C^\lambda =0 ,\qquad \lambda \ge 1.
\end{equation}
For $\lambda_1=2$, both (\ref{25apr5}) and (\ref{25apr3x1}) imply that $f$ has the singularity of the form $1/(k_1^2k_3)$ for both $k_1$ and $k_3$ going to zero. Consistency then requires that the associated prefactors agree, which leads to
\begin{equation}
\label{26apr2}
-6 l C^0 = C^2 C^0.
\end{equation} 
Together with (\ref{26apr1}) it implies that $C^0=0$. Thus, we find that 
\begin{equation}
\label{26apr2x1}
C^\lambda =0 ,\qquad \lambda \ge 0.
\end{equation}

To find the constraints on $\bar C^{-\lambda}$, we consider the consistency conditions for $\lambda_1<0$ and $\lambda_2=0$.
Expression (\ref{4.21}) for $A_{pp}$ in this case is inapplicable, though, we know that $A_{pp}$ contribution in this case anyway vanishes, since $C^0=0$.
Besides that, $A_{pf}$ is also zero, because $C^{\lambda}=0$ for negative $\lambda$. In other words, the only non-vanishing non-homogeneous solution is
\begin{equation}
\label{26apr3}
A^{\l_1 ,0}_{nf}=-3l \bar C^{-\l_1+2}(s_x\sigma_{\bar x}+s_y\sigma_{\bar y})^{-\l_1+2}\frac{(\sigma_x-\sigma_y)s_y}{(\sigma_{\bar x}-\sigma_{\bar y})s_x^{\l_1-1} }(s_x+s_y)^{2\l_1-2}.
\end{equation}
Its only singularity is at $\sigma_{\bar x}\to \sigma_{\bar y}$. Trying to cancel it, we can add a homogeneous solution, which we present in the form
\begin{equation}
\label{26apr4}
A_h^{\l_1,0}= (l^{--})^{\frac{\lambda_1}{2}} (l^{-+})^{\frac{\lambda_1}{2}} \tilde{f}(k_1,k_3) =
s_x^{-\l_1}s_y^{-\l_1}
(\sigma_{\bar x}-\sigma_{\bar y})^{-\l_1}\left( \sigma_{\bar x}\sigma_{y}+\frac{m^2}{2}\right)^{-\lambda_1}
 \tilde{f}(k_1,k_3).
\end{equation}
Note, that we used a different representation for the helicity prefactor. In principle, we could have kept the old expression, but (\ref{26apr4}) seems more convenient as the helicity prefactor it involves is non-singular.  
Then, to cancel the first order pole of (\ref{26apr3}) in $(\sigma_{\bar x}- \sigma_{\bar y})$ one should have $\tilde f(k_1,k_3) \sim \frac{1}{k^{-\lambda_1+1}_3}$.
However, considering that $k_3$ involves a factor of $(\sigma_{ x}- \sigma_{ y})$, this will introduce a pole of order $-\lambda_1+1$ in $(\sigma_{ x}- \sigma_{ y})$, that has nothing to cancel with. In other words, we find that 
\begin{equation}
\label{26apr5}
\bar{C}^{-\lambda}=0, \qquad \lambda <0.
\end{equation}

In summary, (\ref{26apr2}) and (\ref{26apr5}) lead us to the conclusion that a scalar point particle cannot interact with the Poisson chiral higher-spin theory.

\section{Connection to scattering observables}
\label{sec:6}

In the present section, we will interpret the previous discussion in terms of scattering observables. The scattering observables that we will encounter, in addition to the familiar field theory lines involve lines associated with point particles and compute probability amplitudes for scattering of massless fields off a classical point particle. These can be computed either directly by employing the appropriate perturbative schemes featuring point particles from the very beginning, see e. g. \cite{Goldberger:2004jt,Porto:2016pyg,Mogull:2020sak}, 
or accessed via the suitable classical limit of field theory amplitudes replacing fields with point particles, see e. g. \cite{Kosower:2018adc,Maybee:2019jus,Aoude:2021oqj}. We will use the second approach. The relevant limit is the one in which momenta of point particles stay finite, while
the field momenta scale as $\hbar$ when $\hbar \to 0$. Putting it differently, we will consider a limit of field theory amplitudes in which the momenta of scalar particles are much larger than the momenta of massless higher-spin fields. More details on the classical limit can be found in \cite{Kosower:2018adc,Maybee:2019jus,Aoude:2021oqj}.

\subsection{Kinematics}

We start by considering the kinematics of the four-point scattering amplitude involving two external massive scalar lines and two massless higher-spin lines in the classical limit. We will denote the ingoing momenta of point particle lines by $p_1$ and $p_2$, while the ingoing momenta of massless higher-spin fields will be denoted $q_x=i\partial_x$ and $q_y=i\partial_y$. We will assume that all external momenta are on-shell. In particular, 
for the external point particle lines 
\begin{equation}
\label{30apr8}
p_+ \approx -H_p,
\end{equation}
so that
\begin{equation}
\label{30apr9}
p^2 = 2p_-p_+ +2p_x p_{\bar x} \approx -m^2.
\end{equation}
Below, we will always assume that the on-shell conditions are taken into account and $\approx$  will be replaced with the usual equality sign for brevity.

Momentum conservation reads 
\begin{equation}
\label{29may18}
p_1+p_2 + q_x + q_y =0.
\end{equation}
In the classical limit one has
\begin{equation}
\label{31may1}
 q_{\{x,y \}} \sim \hbar \; p_{\{ 1,2\}}, \qquad \hbar \to 0.
\end{equation}
Accordingly, in the limit momentum conservation leads to
\begin{equation}
\label{31may2}
p_1 = -p_2 \equiv p,
\end{equation}
where $p$ should be identified with the point particle momentum used in the preceding part of the paper.

For the given scattering process one can construct the standard set of manifestly Lorentz invariant Mandelstam variables. In particular, one has
\begin{equation}
\label{31may3}
-ip_i \cdot q_x =p_i \cdot \partial_x = - k_{1,i} ,\qquad -ip_i \cdot q_y =p_i \cdot \partial_y = - k_{2,i}.
\end{equation}
Here
\begin{equation}
\label{31may4}
k_{1,i}\equiv k_1\Big|_{p\to p_i}, \qquad
		k_{2,i}\equiv k_2\Big|_{p\to p_i}
\end{equation}
are the same $k_1$ and $k_2$, which we introduced previously (\ref{kkkk}), (\ref{4june1}), except that they depend on $p_1$ or $p_2$ instead of $p$.
As usual, momentum conservation implies that out of four variables in (\ref{31may3}) only two  are independent
\begin{equation}
\label{31may5}
k_{1,1} = k_{2,2}, \qquad k_{1,2} = k_{2,1}.
\end{equation}
One can also construct the last Lorentz invariant Mandelstam variable 
\begin{equation}
\label{29may6}
-q_x\cdot q_y =\partial_x\cdot \partial_y = -k_3,
\end{equation}
which is equal to $ (-p_1\cdot p_2)$ as a consequence of momentum conservation. 

In the  classical limit (\ref{31may1}), we find
\begin{equation}
\label{31may6}
k_{1,1} \to k_1, \qquad k_{1,2} \to - k_1, \qquad k_{2,1} \to k_2, \qquad k_{2,2} \to - k_2.
\end{equation}
Momentum conservation (\ref{31may5}) then leads to
\begin{equation}
\label{31may 7}
k_1+k_2 \equiv k =0,
\end{equation}
thus, giving an alternative perspective on the on-shell condition (\ref{21apr1}).

In summary, we find that a Lorentz invariant  scattering observable, resulting from the classical limit defined above, may depend on two independent Mandelstam variables given by $k_1$ and $k_3$. This result is consistent with (\ref{21apr2}), where we found that Lorentz invariance constrains the point particle light-cone interaction vertex up to an arbitrary function of $k_1$ and $k_3$. In (\ref{21apr2}) we also encountered $l^{\pm \pm}$ factors responsible for changing helicities. Two of these -- $l^{++}$ and $l^{--}$ -- do not depend on the point particle momentum and up to numerical factors are equal to the squares of the spinor products $[12]^2$ and $\langle 12\rangle^2$\footnote{Exact expressions for the off-shell spinor products in terms of the light-cone variables can be found in \cite{Ananth:2012un,Akshay:2014qea,Bengtsson:2016jfk,Bengtsson:2016alt,Ponomarev:2016cwi}.}. The remaining two -- $l^{+-}$ and $l^{-+}$ -- do involve the particle momentum non-trivially and, thus, provide an extension of the spinor-helicity machinery to scattering that features massive point particles. 

\subsection{Exchanges}
In the present section we will give a scattering interpretation of the particular solutions for the point-particle vertices (\ref{4.2}) and (\ref{4.21}).
To be more precise, we will show that the associated contact diagrams cancel contributions of exchanges, thus, resulting in a vanishing and, therefore, manifestly Lorentz invariant amplitude.

In order to do that, let us consider a point particle vertex associated with (\ref{4.1}) for the higher-spin field of non-negative helicity
\begin{equation}
\label{31may8}
S_{p,1} = - \int \frac{dx^+}{p^+} C^\lambda \sigma^{\lambda}_{x} \Phi^\lambda (x).
\end{equation}
Before computing a field-theory amplitude, we need to promote (\ref{31may8}) to a field theory vertex
\begin{equation}
\label{31may9}
S_{f,3} = - i^\lambda \int d^4 x C^\lambda \frac{\bar{\mathbb{P}}^\lambda_{13}}{(\partial_3^+)^\lambda}\varphi(x_1)\varphi(x_2)\Phi^\lambda(x_3)\Big|_{x_i=x},
\end{equation}
where 
\begin{equation}
\label{31may10}
\bar{\mathbb{P}}^\lambda_{13} \equiv \partial_{x_1} \partial_3^+ - \partial_{x_3}\partial_1^+
\end{equation}
and $\varphi$ is a scalar field to be identified with a point particle in the classical limit. 
Vertex (\ref{31may9}) is just the field theory vertex (\ref{19feb1}), (\ref{19feb1x1}), which is integrated by parts as well as has the coupling constant fixed so that in the classical limit it results in (\ref{31may8}). A detailed discussion of the classical limit at the level of action can be found in \cite{Ivanovskiy:2024ads}. Here and in what follows we will assume that helicities are such that cubic vertices are non-vanishing. For example, we will assume that $\lambda$ is even, otherwise (\ref{31may9}) vanishes, which can be seen by using integration by parts. 

Let us now proceed to the computation of an exchange amplitude, which involves a vertex (\ref{31may9}) connected to a field theory vertex (\ref{19feb1}), (\ref{19feb1x1}) with a field theory propagator. The vertex contribution associated with (\ref{31may9}) is
\begin{equation}
\label{29may4}
V_L = -2 i C^{\lambda} \frac{(p_{1x} (\partial_x^++\partial_y^+)-p_1^+(\partial_x+\partial_y))^{\lambda}}{(\partial_x^+ + \partial_y^+)^{\lambda}}.
\end{equation}
Here $i(\partial_x + \partial_y)$ is an exchanged momentum\footnote{Hopefully, writing $q_x = i\partial_x$ and $q_y =i\partial_y$ for momenta of massless fields immediately will not confuse anyone. This allows us to use the notations for various combinations of derivatives and relations between them we gave previously.}, while $2$ is a symmetry factor, which accounts for permutations of two identical fields $\varphi$ in (\ref{31may9}).
Similarly, for the field theory vertex, we have
\begin{equation}
\label{29may3}
V_R = -6 i   C^{\lambda_1,\lambda_2,-\lambda} \frac{(\partial_x \partial_y^+-\partial_y \partial_x^+)^{\lambda_1+\lambda_2-\lambda}}{(\partial_x^+)^{\lambda_1}(\partial_y^+)^{\lambda_2}(-\partial_x^+ - \partial_y^+)^{-\lambda}}.
\end{equation}
Finally, the propagator reads
\begin{equation}
\label{29may5}
P= - \frac{i}{(i\partial_x + i\partial_y)^2}= \frac{i}{2(\partial_x\cdot \partial_y)} = -\frac{i}{2 k_3}.
\end{equation}

Combining all the factors together, we find the exchange amplitude
\begin{equation}
\label{29may7}
\begin{split}
&{\cal A}_{1} = (-6 i) C^{\lambda_1,\lambda_2,-\lambda} \frac{(\partial_x \partial_y^+-\partial_y \partial_x^+)^{\lambda_1+\lambda_2-\lambda}}{(\partial_x^+)^{\lambda_1}(\partial_y^+)^{\lambda_2}(-\partial_x^+ - \partial_y^+)^{-\lambda}}
(-)\frac{i}{2 k_3}\\
& \qquad \qquad  \qquad \qquad\qquad \qquad( -2 i) C^{\lambda} \frac{(p_{1x} (\partial_x^++\partial_y^+)-p_1^+(\partial_x+\partial_y))^{\lambda}}{(\partial_x^+ + \partial_y^+)^{\lambda}}.
\end{split}
\end{equation}
To simplify it, we use
\begin{equation}
\begin{split}
\label{29may8}
p_{1x} (\partial_x^++\partial_y^+)-p_1^+(\partial_x+\partial_y) &= p_1^+ (s_x\sigma_x +s_y \sigma_y),\\
\bar{\mathbb{P}}_{xy}\equiv \partial_x \partial_y^+ - \partial_y \partial_x^+& = - \frac{\partial_x \partial_y}{p_1^+}(\sigma_x - \sigma_y).
\end{split}
\end{equation}
Eventually, in the classical limit we just replace $p_1$ with $p$, which gives
\begin{equation}
\label{29may10}
{\cal A}_{1} \to 6 i C^{\lambda_1,\lambda_2,-\lambda} C^{\lambda}  \frac{1}{k_3} \frac{(\sigma_x-\sigma_y)^{\lambda_1+\lambda_2-\lambda} (s_x\sigma_x  + s_y \sigma_y)^\lambda}{s_x^{\lambda-\lambda_2}s_y^{\lambda-\lambda_1}}, \qquad \hbar \to 0.
\end{equation}
To arrive at (\ref{29may10}) we assumed that $\lambda_1+\lambda_2$ is even, otherwise, the field theory vertex (\ref{29may3}) with even $\lambda$ vanishes.
Keeping in mind the symmetry factor of $2!2!$ accounting for permutations of two scalar and two higher-spin lines, it is not hard to see that the contact contribution associated with the particular solution
(\ref{4.2}) cancels (\ref{29may10}) exactly. 

Next, we proceed to the exchanges with point particles\footnote{By this we only mean that the respective Feynman diagrams feature scalar fields on the internal lines and these will become point particles in the classical limit. This does not imply that point particles are created in the scattering process or that these mediate any  interactions. }. There are two exchange diagrams of this type. For the first one, we have the following vertex
\begin{equation}
\label{29may12}
V_{L} = (-2 i )C^{\lambda_1} \left( \frac{p_{1x}\partial_x^+ - p_1^+ \partial_x}{\partial_x^+} \right)^{\lambda_1} , 
\qquad
V_{R} = (-2 i )C^{\lambda_2} \left( \frac{p_{2x}\partial_y^+ - p_2^+ \partial_y}{\partial_y^+} \right)^{\lambda_2}
\end{equation}
 and propagator factors
\begin{equation}
\label{29may14}
P = -\frac{i}{(p_1+ i \partial_x)^2 + m^2} = - \frac{i}{2i p_1 \cdot \partial_x} = \frac{1}{2k_{1,1}}.
\end{equation}
Combining them together, we find
\begin{equation}
\label{29may17}
{\cal A}_2 =  - 2  C^{\lambda_1} C^{\lambda_2} \frac{1}{k_{1,1}} 
 \left( \frac{p_{1x}\partial_x^+ - p_1^+ \partial_x}{\partial_x^+} \right)^{\lambda_1}
 \left( \frac{p_{2x}\partial_y^+ - p_2^+ \partial_y}{\partial_y^+} \right)^{\lambda_2} .
\end{equation}

As we will see below, in the classical limit (\ref{29may17}) cancels the classical limit of the exchange diagram with a point particle exchanged in the remaining channel. 
To find a non-vanishing expression, we have to expand the exchanges in small momenta and keep the subleading terms. In order to proceed, we will use $p_1 =p$ in the classical limit, while $p_2$ will be expressed in terms of $p_1$ and small momenta using momentum conservation. We then find 
\begin{equation}
\label{29may19}
\frac{p_{2x}\partial_y^+ - p_2^+ \partial_y}{\partial_y^+}= -\sigma_y - i \frac{\bar{\mathbb{P}}_{xy}}{\partial_y^+}
\end{equation}
and (\ref{29may17}) acquires the form
\begin{equation}
\label{29may20}
{\cal A}_2 =  - 2  C^{\lambda_1} C^{\lambda_2} \frac{1}{k_{1}} 
\sigma_x^{\lambda_1}
 \left( -\sigma_y - i \frac{\bar{\mathbb{P}}_{xy}}{\partial_y^+} \right)^{\lambda_2} .
\end{equation}

We now move on to the point particle exchange in the remaining channel. It is given by
\begin{equation}
\label{29may21}
\begin{split}
{\cal A}_3 = (-2 i )C^{\lambda_2} \left( \frac{p_{1x}\partial_y^+ - p_1^+ \partial_y}{\partial_y^+} \right)^{\lambda_2} 
\left(-\frac{1}{2p_1\cdot \partial_y} \right)
(-2 i )C^{\lambda_1} \left( \frac{p_{2x}\partial_x^+ - p_2^+ \partial_x}{\partial_x^+} \right)^{\lambda_1} .
\end{split}
\end{equation}
Keeping the subleading terms, we have
\begin{equation}
\label{29may23}
p_1\cdot \partial_y = -p_1\cdot  \partial_x - i \partial_x \cdot\partial_y
  = k_{1} + i k_3.
  \end{equation}
Then, for (\ref{29may21}) we find 
\begin{equation}
\label{29may24}
{\cal A}_3 =  - 2  C^{\lambda_1} C^{\lambda_2} \frac{1}{ - k_{1}- i k_3} 
 \left( \frac{p_{1x}\partial_y^+ - p_1^+ \partial_y}{\partial_y^+} \right)^{\lambda_2}
 \left( \frac{p_{2x}\partial_x^+ - p_2^+ \partial_x}{\partial_x^+} \right)^{\lambda_1}.
 \end{equation}
Proceeding with the expressions in the brackets in the same manner as before, we obtain
\begin{equation}
\label{29may24x1}
{\cal A}_3 =   2 C^{\lambda_1} C^{\lambda_2} \frac{1}{  k_{1}+ i k_3} 
\sigma_y^{\lambda_2}
 \left(- \sigma_x + i \frac{\bar{\mathbb{P}}_{xy}}{\partial_y^+} \right)^{\lambda_1}.
 \end{equation}

Summing the two contributions (\ref{29may20}) and (\ref{29may24x1}), we find 
\begin{equation}
\label{29may26}
\begin{split}
{\cal A}_2+{\cal A}_3 = (-2)C^{\lambda_1} C^{\lambda_2}
\left[\frac{1}{k_{1}}\sigma_x^{\lambda_1}
 \left( \sigma_y + i \frac{\bar{\mathbb{P}}_{xy}}{\partial_y^+} \right)^{\lambda_2}
 -
 \frac{1}{  k_{1}+ i k_3} 
\sigma_y^{\lambda_2}
 \left( \sigma_x - i \frac{\bar{\mathbb{P}}_{xy}}{\partial_y^+} \right)^{\lambda_1}
 \right].
 \end{split}
\end{equation}
Here again we assumed that both $\lambda_1$ and $\lambda_2$ are even, otherwise, the diagram vanishes.
In the classical limit
\begin{equation}
\label{31may11}
\frac{k_3}{k_1}\sim \hbar, \qquad \frac{\bar{\mathbb{P}}_{xy}}{\partial_y^+\sigma_y} \sim \hbar, \qquad \frac{\bar{\mathbb{P}}_{xy}}{\partial_x^+\sigma_x} \sim \hbar.
\end{equation}
By expanding (\ref{29may26}) in small variables (\ref{31may11}), we find that the first non-trivial order is the subleading one, for which we have
\begin{equation}
\label{29may27}
\begin{split}
{\cal A}_2+{\cal A}_3 \to 2i  C^{\lambda_1} C^{\lambda_2} \sigma_x^{\lambda_1} \sigma_y^{\lambda_2}
\left[\frac{1}{k_1} \frac{\sigma_x - \sigma_y}{\sigma_x \sigma_y}(\sigma_x s_x \lambda_2 + \sigma_y s_y \lambda_1) - \frac{k_3}{k_1^2}
 \right], \qquad \hbar \to 0.
 \end{split}
\end{equation}
It is not hard to see that the contact contribution, resulting from the particular solution (\ref{4.21}) cancels (\ref{29may27}) exactly.

In summary, we showed that the point particle vertices (\ref{4.2}) and (\ref{4.21}) cancel the respective exchange diagrams, resulting in a vanishing scattering observable. The same holds for the contribution  (\ref{4.22}) corresponding to the exchange with a field of negative helicity. Then, the fact that (\ref{4.2}), (\ref{4.22}) and  (\ref{4.21}) provide a particular solution for the Hamiltonian translates into a simple statement that the vanishing scattering amplitude is Lorentz invariant. By adding to the particular solution of the non-homogeneous consistency conditions non-trivial solutions of homogeneous equations we arrive at non-trivial amplitudes. In these terms the analysis of section \ref{sec:5}
amounts to studying whether Lorentz invariant amplitudes can accommodate singularities contributed by exchanges. 

Finally, let us remark that the second order pole in (\ref{29may27}), though, may seem unusual from the field theory perspective,  is typical for worldline scattering observables, see e. g. \cite{Shen:2018ebu,Shi:2021qsb}. Above we found that it naturally arises as a result of collision of two simple poles associated with exchanges in different channels in the classical limit.

\section{Conclusion}
\label{sec:7}

In the present paper we studied the light-cone consistency conditions for interactions of a massive scalar point particle with fields of chiral higher-spin theories at the second order in the coupling constant. We found that these interactions are prohibited, thus, extending our previous result \cite{Ivanovskiy:2025kok} to massive point particles. 
As a consistency check of our approach, we also studied interactions of a massive point particle with self-dual gravity and found that consistent interactions in this case do exist, reproducing the well-known covariant result. As explained in more detail in our previous papers \cite{Ivanovskiy:2023aay,Ivanovskiy:2025kok}, the existence of a consistent coupling of a scalar point particle to massless higher-spin fields is closely related to the existence of a higher-spin extension of Riemannian geometry. 
In the present paper we  completed the analysis of consistent interactions between scalar  point particles and chiral higher-spin fields  showing that these do not exist, which, considering that chiral higher-spin theories form inevitable closed subsectors of interacting massless higher-spin theories in flat space, also implies, that extensions of Riemannian geometry to space-times with non-trivial higher-spin fields are not possible. 

In the course of our work we faced the necessity to solve the light-cone consistency conditions at the second order in the coupling constant. This is a rather complicated problem and it is not entirely clear how to approach it systematically. In the case when interactions of massless fields in four-dimensional Minkowski space are considered, a systematic approach towards solving the light-cone consistency conditions at all orders was developed in  \cite{Ponomarev:2016cwi}. It allows one to replace a more tedious analysis in the light-cone formalism with much simpler manipulations in terms of spinor-helicity amplitudes.

When solving the light-cone consistency conditions for interactions of point particles with higher-spin fields in the present paper we found that the scattering observables naturally emerge as well. The relevant scattering observables involve point particles and can be obtained as the classical limit of the standard scattering amplitudes,
see e. g. \cite{Kosower:2018adc,Maybee:2019jus,Aoude:2021oqj}. To be more precise, we found that the general solution for the Hamiltonian at the given order is given as the sum of a singular particular solution of the non-homogeneous consistency conditions and the general solution of the homogeneous ones. A simple computation shows that the particular solution results in the vanishing scattering. It is manifestly Lorentz invariant, so it is to be expected that the associated Hamiltonian solves the light-cone consistency conditions.
 In turn, homogeneous solutions correspond to non-trivial Lorentz invariant scattering observables. 

Insisting on locality of the Hamiltonian, we tried to cancel the singularities contributed by the particular solution by the appropriate choice of the solution of the homogeneous consistency conditions. In amplitude terms this translates into the requirement that singularities of amplitudes should match those of exchanges. Thus, our procedure of imposing locality can be regarded as a version of the on-shell methods, see e. g. \cite{Britto:2004ap,Britto:2005fq,Benincasa:2007xk,Benincasa:2011pg,Arkani-Hamed:2017jhn}, applied to worldline theories. We found that the singularities of the particular solution for the Hamiltonian cannot be cancelled, thus, local and Lorentz invariant interactions of scalar point particles with chiral higher-spin fields do not exist. 

Overall, it is, certainly, not surprising that consistent -- in the sense of the light-cone deformation procedure -- Hamiltonians lead to consistent scattering and vice versa. 
Moreover, this connection should be universal in the sense that it should not rely on the  details of particular theories. 
At the same time, within the light-cone formalism such a connection is not obvious, as can be seen from the respective analysis  for massless fields \cite{Ponomarev:2016cwi}. Our results confirm the expectation that this connection is also valid for worldline theories and that it can be helpful for solving the light-cone consistency conditions in this case.  It would be interesting to establish the connection between the light-cone consistency conditions and the consistency conditions on the associated scattering in the universal form in future. 
This would not only allow to facilitate the light-cone formalism, but it can also enrich the amplitude methods. In particular, in contrast to amplitudes, which are on-shell observables, the light-cone approach operates at the level of action, therefore, it provides an off-shell extension of amplitude techniques\footnote{For massless fields in 4d Minkowski space the relevant extension appeared in \cite{Bardeen:1995gk,Cangemi:1996rx,Cachazo:2004kj}.}. Besides that, unlike typical amplitude methods, the light-cone approach does not rely on manifest Lorentz invariance and it may, in principle, result in new efficient representations for scattering observables,  such as the spinor-helicity formalism for massless fields in flat space-time\footnote{For various extensions of the spinor-helicity formalism, see \cite{Cheung:2009dc,Maldacena:2011nz,Conde:2016vxs,Bandos:2017eof,Nagaraj:2018nxq,Basile:2024ydc} and references therein.}.

\acknowledgments

We would like to thank A. Ochirov for helpful correspondence and E. Skvortsov for comments on the manuscript.

\bibliography{pp}
\bibliographystyle{JHEP}

\end{document}